\documentclass[apj]{emulateapj}
\usepackage{apjfonts}
\usepackage{graphicx}
\usepackage{color}
\usepackage{amsbsy}
\usepackage{mathrsfs}
\usepackage{mathpazo,bm}
\usepackage{amsmath}
\usepackage{multirow}

\newlength{\hfwidth}
\newlength{\hfwidthsingle}
\renewcommand{\v}[1]{{\boldsymbol{#1}}} 

\newcommand{\del}{\v{\nabla}}
\newcommand{\grad}{\del}
\newcommand{\Div}{\del\cdot}

\newcommand{\ttimes}[1]{10^{#1}}

\newcommand{\Eq}[1]{Eq. (\ref{#1})}

\newcommand{\Eqs}[2]{Eqs. (\ref{#1}) and~(\ref{#2})}

\newcommand{\eq}[1]{\Eq{#1}}

\newcommand{\Figure}[1]{Figure~\ref{#1}}
\newcommand{\Fig}[1]{Fig.~\ref{#1}}
\newcommand{\fig}[1]{\Fig{#1}}

\newcommand{\sect}[1]{Sect.~\ref{#1}}
\newcommand{\capsect}[1]{Section~\ref{#1}}
\newcommand{\Tab}[1]{Table \ref{#1}}
\newcommand{\tab}[1]{\Tab{#1}}

\newcommand{\mearth}{$M_{\earth}$ }

\shorttitle{Orbital migration in evolutionary models}
\shortauthors{Horn et al.}

\begin{document}

\title{Orbital migration of interacting low-mass planets \\
       in evolutionary radiative turbulent models}

\author{Brandon Horn   \altaffilmark{1},
Wladimir Lyra          \altaffilmark{2,3,4},\\
Mordecai-Mark Mac Low  \altaffilmark{1,2,5}, \\ and \\
Zsolt S\'andor         \altaffilmark{5,6}}

\altaffiltext{1}{Department of Astronomy, Columbia University, 
                 550 West 120th St, New York, NY 10027, USA; 
                 bhorn@astro.columbia.edu}
\altaffiltext{2}{Department of Astrophysics, American Museum of
                 Natural History, 79th Street at Central Park 
                 West, New York, NY 10024, USA; 
                 wlyra@amnh.org, mordecai@amnh.org}
\altaffiltext{3}{Jet Propulsion Laboratory, 4800 Oak Grove Drive, 
                 Pasadena, CA 91109, USA}
\altaffiltext{4}{NASA Carl Sagan Fellow}
\altaffiltext{5}{Max-Planck-Institut f\"ur Astronomie, K\"onigstuhl 
                 17, 69117, Heidelberg, Germany} 
\altaffiltext{6}{Institut f\"ur Astro und Teilchenphysik, 
                 Leopold Franzens Universit\"at Innsbruck, 
                 Technikerstrasse 25/8, A-6020 Innsbruck, Austria;
                 zsolt.sandor@uibk.ac.at} 

\begin{abstract} {
The torques exerted by a locally isothermal disk on an embedded planet 
lead to rapid inward migration. Recent work has shown that modeling the 
thermodynamics without the assumption of local isothermality reveals regions 
where the net torque on an embedded planet is positive, leading to outward 
migration of the planet. When a region with negative torque lies directly 
exterior to this, planets in the inner region migrate outwards and planets 
in the outer region migrate inwards, converging where the torque is zero. 
We incorporate the torques from an evolving non-isothermal disk into an $N$-body
simulation to examine the behavior of planets or planetary embryos interacting 
in the convergence zone. We find that mutual interactions do not eject 
objects from the convergence zone.  Small numbers of objects in a laminar 
disk settle into near resonant orbits that remain stable over the 10 Myr 
periods that we examine. However, either or both increasing the number of 
planets or including a correlated, stochastic force to represent turbulence 
drives orbit crossings and mergers in the convergence zone. These processes 
can build gas giant cores with masses of order ten Earth masses from 
sub-Earth mass embryos in 2--3 Myr.}  
\end{abstract}

\section{Introduction}
\label{sect:introduction}

The details of how giant planets form in protoplanetary disks remain poorly 
understood. The core accretion model developed by \citet{Pollack} relies on 
the rapid growth of a gas giant through runaway accretion of gas onto a solid 
planetary core that grows massive enough to gravitationally capture the gas.
Such cores form through binary collisions of planetary embryos---the bodies 
formed during oligarchic growth, with masses ranging from 0.1 $M_{\earth}$ to 
a few Earth masses. Since the accretion must necessarily occur in a gas-rich 
disk, the planetary core must be formed before the gas disk dissipates. The 
mass required to begin rapidly accreting gas is determined by the aspect 
ratio of the disk, the mass of the star, and the luminosity of the 
core that is provided by planetesimal accretion and can prevent collapse of 
the accreting gas. For a solar mass star, a thin 
disk with an aspect ratio of 0.05, and a surface density of planetesimals 
of 10 g\,cm$^{-2}$ at 5\,AU, the core must be around ten Earth masses 
to begin runaway accretion. The lifetime of the gas disk is on the order of 
10 Myr, and the core must form before the disk dissipates. 

Planetary embryos are massive enough to not be affected by the gas drag 
forces that dominate the motion of smaller bodies moving through the 
sub-Keplerian gas disk. The mass of the planetary embryos, however, induces 
spiral Lindblad resonances in the gas disk. The asymmetry in the size of the 
inner and outer Lindblad resonances generates a net negative torque from the 
disk that, in the absence of other effects, robs the planet of angular 
momentum, leading to inward Type I migration and ultimately infall into the central 
star. A linear treatment of the torque from isothermal gas in the co-rotating 
region of the planetary embryo shows that it provides a positive torque 
arising from the presence of a vortensity gradient that nearly offsets the 
net Lindblad torque, but a small net negative torque remains.

The behavior of multiple planetary embryos embedded in a gas disk and 
undergoing inward migration due to Lindblad and linear co-rotational torques 
was explored by \citet{Cresswell}. They found that as bodies with different 
masses migrate inwards, differential migration leads to mergers, forming 
bodies with masses in the planetary core range. These  simulations use large 
embryos--with masses of at least 2 $M_{\earth}$ and up to planetary core 
mass--and simulate migration in a laminar, isothermal disk.

In addition to the perturbations to the gas disk caused by the presence of the 
planetary embryos, a partially ionized disk with a subthermal magnetic field 
can develop turbulence through the magnetorotational instability
\citep[MRI,][]{Balbus91}.  The resulting random overdensities in the gas can lead the 
planetary embryo to undergo a random walk, possibly avoiding infall for some 
of the bodies \citep{Nelson}. While this is an effective way to
prevent catastrophic infall for planetesimal-sized objects with radii 
of $<1000$~km \citep{Yang09,Yang11}, it does not provide a clear mechanism 
for building large planetary cores. 

Planetary embryos undergoing inward Type~I migration can be trapped in pressure 
ridges that may form in the presence of a dead-zone boundary or within an 
anti-cyclonic vortex. The positive gradient in 
surface density in these regions leads to formation of a convergence zone, 
where migration halts. The dynamics of multiple planetary embryos 
undergoing Type~I migration in the 
proximity of a pressure ridge convergence zone have been investigated by 
\citet{Morbidelli08}, who found that planets collide until only a few 
remain. \citet{Sandor} showed that 
such pressure traps can lead to growth from planetary embryos to planetary 
cores through continual generation of bodies that migrate into the pressure
ridge convergence zone. 

One possible means of avoiding catastrophic infall that could lead to rapid 
formation of planetary cores was found by \citet{Paard2006}, who explored the 
interactions of planets with a gas disk, in a model in which they did not 
make the locally isothermal approximation, but rather directly modeled the 
effects of radiative transfer through optically thick gas with an adiabatic 
equation of state. They found that the co-rotation torque arising from the 
gas undergoing horseshoe turns near the planet showed a strong positive 
contribution when the isothermal approximation was relaxed. This was later 
shown to be a result of the presence of an entropy gradient across the 
corotation region \citep{Paard2008,Baruteau-Masset}.
A large radial entropy gradient increases the positive co-rotation torque 
to the point that it cancels out the negative net Lindblad torque and 
can even reverse the direction of migration. This understanding was 
developed further in \citet{Paard2009}. A prescription for the nonlinear 
horseshoe torque was given by \citet{Paard2010}, allowing the total torque 
on a planet undergoing Type~I migration to be calculated, given the local 
temperature and surface density gradients. The migration of a single planet 
embedded in an evolving disk was then studied by \citet[][hereafter 
LPM10]{Lyra}, who found that planets migrate towards convergence zones, where the 
net torque is zero. Planets located exterior to these orbits experience a 
negative torque, while planets located interior experience a positive torque. 
Upon reaching the zero-torque region, the planets migrate inwards with the 
disk until the disk dissipates, thus avoiding collision with the star.

In this paper, we extend the work of LPM10 by following the
interactions of multiple planets converging towards any zero-torque
radius.  We use the disk evolution model from LPM10, for a
non-isothermal one-dimensional disk with
photoevaporation and viscous diffusion.  The radial profiles of
temperature and surface density from the disk evolution model are used
to determine the torque on planetary embryos in an $N$-body
simulation, following the formula for non-isothermal, unsaturated,
horseshoe torque of \citet{Paard2010}. An optimized version of the 
Bulirsch-Stoer $N$-body code used in \citet{Sandor} is employed
to integrate the orbits of the planets under the influence of the
torque from the gas disk, as well as eccentricity and inclination
damping following \citet{Cresswell}. The code was modified to include
the effect of turbulence on the planetary motion, using the
prescription described by \citet{Laughlin}. 

We start in \sect{sect:method} by describing the disk model, the prescriptions for the 
torque, eccentricity and inclination damping, the turbulence model, and how 
they are included in the $N$-body code. In \sect{sect:ic}, we describe the initial 
conditions of the planets in the simulations as well as the parameters 
describing the different gas disks that were used. \capsect{sect:results} presents the 
results of the simulations and \sect{sect:caveats} includes a discussion of the results. 
A study tackling a similar problem was made public after initial
  submission of our paper by 
Hellary \& Nelson (2012, hereafter HN12). We discuss the differences between 
our work and theirs in \sect{sect:hn12}. A summary and conclusions 
are presented in \sect{sect:conclusions}. 

\begin{figure}
  \begin{center}
    \resizebox{\columnwidth}{!}{\includegraphics{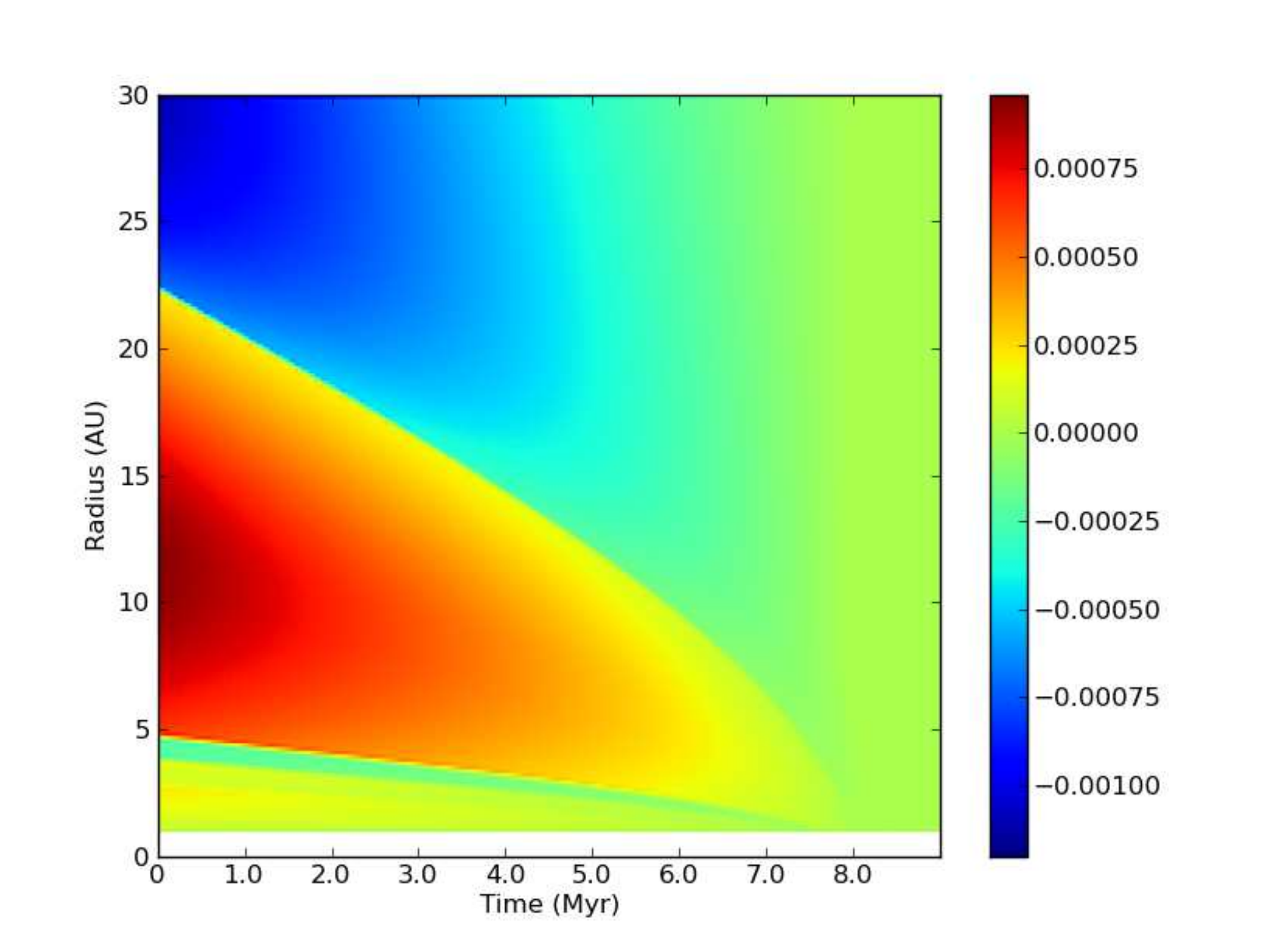}}
  \end{center}
\caption[]{The dimensionless migration torque and how it evolves through
time. The region of interest in these simulations is the transition
from negative torque - inward migration - to positive torque that is 
initially located around 22 AU. As the disk evolves, this convergence 
zone moves inwards. The torque shown in this figure is an interpolation 
between the adiabatic and isothermal torques, based on the opacity at a 
given radius. To find the torque acting on a body at a given location, 
this dimensionless torque must be multiplied by $\Gamma_0$, from 
\eq{eq:scale-torque}.}
 \label{fig:torqueProf}
\end{figure}

\section{Method}
\label{sect:method}
Our simulation can be viewed as a modified $N$-body simulation, which 
incorporates the migration torques acting on planetary embryos,
a stochastic force representing turbulence in the 
gas disk, and eccentricity and inclination damping from the gas 
disk, as well as the disk evolution due to viscosity and 
photoevaporation. The migration torques depend on temperature and 
surface density profiles from one-dimensional simulations of the 
evolution of the gas disk, and the force from turbulence is modeled 
using a time-dependent potential. Each 
component of the simulation is described in more detail in this section.

\begin{deluxetable*}{lcccccccccc}
  \tabletypesize{\tiny}
  \tablecaption{Run Parameters}
  \tablewidth{0pt}
  \tablehead{
    \colhead{Run}                &
    \colhead{$M_p$}              &
    \colhead{$\mu_{\rm mass}$}   & 
    \colhead{$\sigma_{\rm mass}$}&
    \colhead{Combine}            & 
    \colhead{$N_p$}              &
    \colhead{$N_{mHR}$}          & 
    \colhead{$M_{\rm tot}$}      &
    \colhead{Swarm}              &
    \colhead{$T_{\rm form}$}     &
    \colhead{$M_{\rm core}$}\\
    \colhead{(1)} & \colhead{(2)} & \colhead{(3)} & \colhead{(4)} & \colhead{(5)}  & 
    \colhead{(6)} & \colhead{(7)} & \colhead{(8)} & \colhead{(9)} & \colhead{(10)} & 
    \colhead{(11)}}
  \startdata
H1 (Fiducial) & 0.75 & --   & --  & --      & 23 & 3  & 17.25 & No  &  1.29  & 10.5       \\ 
H2            & 0.5  & --   & --  & --      & 27 & 3  & 13.5  & No  &  2.23  & 8.0        \\ 
H3            & 1.00 & --   & --  & --      & 21 & 3  & 21.0  & No  &  0.43  & 13.0       \\ 
H4            & 1.25 & --   & --  & --      & 20 & 3  & 25.0  & No  &  0.21  & 10.0       \\ 
H5            & 1.5  & --   & --  & --      & 19 & 3  & 29.5  & No  &  0.15  & 10.5       \\\hline 
G1            & --   & 1.0  & 0.3 & --      & 21 & 3  & 20.9  & No  &  0.27  & 11.1       \\ 
G2            & --   & 1.0  & 0.3 & --      & 11 & 6  & 11.7  & No  &  0.34  & 7.1        \\ 
G3            & --   & 0.75 & 0.3 & --      & 24 & 3  & 18.7  & No  &  2.38  & 10.2       \\ 
G4            & --   & 0.75 & 0.3 & --      & 13 & 6  & 9.8   & No  &  1.49  & 9.3        \\ 
G5            & --   & 0.5  & 0.2 & --      & 27 & 3  & 13.9  & No  &  3.18  & 9.3        \\ 
G6            & --   & 0.5  & 0.2 & --      & 14 & 6  & 6.2   & No  &  1.38  & 4.4        \\\hline 
G7            & --   & --   & --  & G2$+$G6 & 25 & 6  & 17.9  & No  &  1.36  & 12.0       \\ 
G8            & --   & --   & --  & G2$+$G4 & 24 & 6  & 21.5  & No  &  0.95  & 10.5       \\ 
G9            & --   & --   & --  & G4$+$G6 & 27 & 6  & 16.0  & No  &  2.64  & 9.4        \\\hline 
Sw1           & 0.25 & --   & --  & --      & 25 & 4  & 6.25  & Yes &  4.60  & 2.5        \\ 
Sw2           & 0.75 & --   & --  & --      & 12 & 6  & 9     & Yes &  4.41  & 8.5        \\ 
Sw3           & 1.0  & --   & --  & --      & 11 & 6  & 11    & Yes &  0.68  & 8.5        \\ 
Sw4           & 1.0  & --   & --  & --      & 7  & 10 & 7     & Yes &  1.42  & 9.0        \\\hline 
SD1           & 0.75 & --   & --  & --      & 23 & -- & 17.25 & No  &  0.98  & 10.5       \\
SD2           & 1.0  & --   & --  & --      & 17 & -- & 17.0  & No  &  0.267 & 7.0 \& 5.0 \\ 
SD3           & --   & 0.75 & 0.3 & --      & 21 & -- & 17.26 & No  &  1.03  & 10.2       \\
  \enddata
  \tablecomments{
    Col.\ (1): Name of run. 
    Col.\ (2): Initial mass of bodies, in $M_\earth$. 
    Col.\ (3): Mean of Gaussian distribution for initial masses, in $M_\earth$. 
    Col.\ (4): Standard deviation of Gaussian distribution for initial masses, in $M_\earth$. 
    Col.\ (5): Combined initial sample. 
    Col.\ (6): Number of initial planets. 
    Col.\ (7): Initial separation between planets, in mutual Hill radii. 
    Col.\ (8): Total initial mass. 
    Col.\ (9): Presence of background swarm of small (0.25 $M_\earth$) planets. 
    Col.\ (10): Formation time (Myr). 
    Col.\ (11): Mass of the core formed, in $M_\earth$.}
  \label{table:runs}
\end{deluxetable*}

\subsection{Disk Evolution}
\label{sect:non-iso-disk}

The gas disk evolves via viscous diffusion and photoevaporation. The 
evolution of the gas surface density due to viscosity is given by \citep{L-B}
\begin{equation}
  \frac {\partial \varSigma}{\partial t} = 
  \frac{3}{r} \frac{\partial}{\partial r} \left[r^{1/2} 
  \frac{\partial \varSigma \nu r^{1/2}}{\partial r} \right],
\end{equation}
where $r$ is the distance from the central star, and 
$\nu$ is the kinematic viscosity. Following \citet{Shakura} we use
\begin{equation}
  \nu = \alpha c_s^2 / \varOmega
\end{equation}
where $\alpha$ is a dimensionless parameter. The Keplerian angular velocity
is
\begin{equation} 
  \varOmega = (G M_{\star}/r^3)^{1/2}, 
\end{equation} 
where $M_{\star}$ is the mass of the central star, and $c_s$
is the local sound speed. Our simulation uses a model based 
on the results of \citet{Hollenbach} 
to specify the rate of surface density loss due to photoevaporation. 
This occurs only beyond a radius $r_g$ where ionized disk gas has 
a sound speed equal to the escape velocity. In this regime
\begin{equation}
\dot{\varSigma}_{\rm w} =  \frac{\dot{M}_{\rm w}\, r_g^{\,1.5}}
         {2\pi \left(r_{\rm ext}-r_g\right)r^{2.5}},
\end{equation}
while $\dot{\varSigma}_{\rm w} = 0$ within.
The rate of photoevaporation is set by 
the wind mass loss rate $\dot{M}_{\rm w}$, the 
escape radius $r_g$, and the external boundary 
of our simulation $r_{\rm ext}$. As the gas disk evolves, 
we record temperature and surface density profiles every 10$^3$ yr. 

For temperature evolution, we use a model without shock heating
\citep{Nakamoto} that balances viscous heating, background
radiation, and radiative cooling to derive an emerging flux
\begin{equation}
2\sigma{T^4}=\tau_{\rm eff}\left(\frac{9}{4}\varSigma\nu\varOmega^2\right)+2\sigma{T_b^4},\label{eq:temperature}
\end{equation}
where $T$ and $T_b$ are the midplane and background temperatures,
and $\sigma$ the Stefan-Boltzmann
constant. We take 
the effective optical depth at 
the midplane \citep{Hubeny,Kley08}
\begin{equation}
\tau_{\rm eff}=\frac{3\tau}{8}+\frac{\sqrt{3}}{4}+\frac{1}{4\tau}.\label{eq:taueff}
\end{equation}
The optical depth is $\tau=\kappa\varSigma/2$, and 
the opacities $\kappa$ are taken from \citet{Bell}. We assume that although 
dust growth and planet formation lock away refractory material, 
fragmentation efficiently replenishes small grains, keeping 
the disks opaque during their evolution \citep{Birnstiel}.

The gas simulations are conducted using the same numerical
algorithm as LPM10, which evolves the gas disk in one dimension over the lifetime
of the disk. As a numerical solver, we used the {\sc Pencil Code}
\footnote{The code,
  including improvements done for the present work, is publicly
  available under a GNU open source license and can be downloaded at
  http://www.nordita.org/software/pencil-code}, which integrates
  the equations of evolution with sixth order spatial derivatives, and
  a third order Runge-Kutta time integrator. Boundary conditions are
taken as outflow. Because the optical depth depends on temperature, we
solve \eq{eq:temperature} with a Newton-Raphson root-finding algorithm
(using 0.01\,K precision). To save on computational time, temperatures
are pre-computed as a function of $\varSigma$ and $\varOmega$, stored
in the memory as look-up tables, and retrieved in run time via
bilinear interpolation.

The reduction to one dimension assumes azimuthal symmetry and that the
disk can be vertically integrated without significantly changing the
characteristics of the disk. The azimuthal asymmetry will be
reintroduced into the simulation via our migration and turbulence
prescriptions, and the scale-height parameterizes the effects of a
finite disk height on the magnitude of the turbulence as well as
setting the timescales for eccentricity and inclination damping.

The profiles from the gas simulations are used to determine the local
temperature and gas surface density in the $N$-body code. The
migration torques and turbulent forces acting on the planets are
proportional to the local surface density 
$\varSigma$ and temperature $T$. The local
values of $\varSigma$ and $T$ at any given time and
 position are found by linearly interpolating from the two closest
times, and then again between the two nearest radial grid points.

Several quantities are indirectly dependent on $T$ by way of the 
local sound speed, 
\begin{equation}
  c_s = \sqrt{\frac{\gamma\,kT}{\mu \, m_{\rm H}}},
  \label{eq:sound-speed}
\end{equation}
where $\gamma$ is the adiabatic coefficient of the gas, 
set to 7/5 for diatomic molecules, $k$ is Boltzmann's constant, 
$\mu$ is the mean molecular weight of the gas, set to 
2.4 for a 5:1 H$_2$-He mixture, and $m_{\rm H}$ is the mass of 
a hydrogen atom. The sound speed is also used to determine 
the aspect ratio of the disk
\begin{equation} \label{eq:aspect}
   h= \frac{H}{r}= \frac{c_s}{v_K}
\end{equation}
where $H$ is the scale height of the disk, and the Keplerian 
velocity $v_K=(GM_{\star}/r)^{1/2}$. Unless otherwise noted, 
$\varSigma$, $c_s$, $H$ and $h$ are time and radius dependent, with 
values derived from the gas evolution model.

The local radial gradients of temperature and surface density
needed for the torque calculation are calculated for the entire 
disk during its evolution
using a sixth-order, finite-difference method, as is done in
the {\sc Pencil Code}.
The gradients are recorded with the density and temperature
profiles. During the $N$-body simulation, the same linear
interpolation method is used for the gradients as for the original
quantities.

We ignore gap opening by massive planets, since that does not occur 
for planets with masses less than 10--20~$M_{\earth}$ \citep{Crida}, 
and the goal of this simulation 
is to examine whether planetary embryos can coalesce into planets with 
masses approaching this regime. Therefore, we can simplify the
simulations by evolving  
the gas disk for its entire lifetime before beginning the $N$-body 
simulations. 

\subsection{Prescription for migration}
\label{subsec:torque}
Planets in a gas disk induce spiral Lindblad resonances in the disk,
as well as driving the gas in the co-orbital region into horseshoe
orbits. The net torque on the
planet from the Lindblad resonances can be calculated analytically, as in
\citet{Ward}. In isothermal, non-magnetized disks, the full net 
torque (Lindblad and co-rotational) can be
calculated semi-analytically \citep{Tanaka}. The Lindblad resonances 
impart a negative torque to the planet, leading to
the well known effect  
of rapid inward migration. Hydrodynamical simulations of a planet with 
mass on the
order of $M_{\earth}$ embedded in an isothermal disk have confirmed
the predictions from linear theory \citep{Bate,D'Angelo}. 

\citet{Paard2006} investigated the migration of a low-mass planet in
an adiabatic, hydrodynamic simulation including
radiative transfer for cooling radiation, and showed that outward migration
can occur in such a non-isothermal disk
due to positive torques exerted by the co-rotation region under
some circumstances. Later investigations showed
that this departure from isothermal Type~I migration is caused by the
radial entropy gradient of the disk
\citep{Paard2008,Baruteau-Masset}. A 
prescription incorporating the effects of non-isothermal
co-rotation torques was developed by \citet{Paard2010}.

The prescription allows us to calculate the torque acting on a
low-mass planet embedded in a gas disk as a function of the radial
gradient of the disk's temperature and density in the region near the
planet. Defining the gradients as
\begin{equation}
 \eta = \frac{\partial \ln \varSigma}{\partial \ln r};~~\beta = \frac{\partial \ln T}{\partial \ln r};~~\zeta = \beta - \left(\gamma - 1\right)\eta,
\end{equation}
\citet{Paard2010} arrived at the following equation for the total 
unsaturated torque on the planet
\begin{equation}
 \gamma \Gamma_{\rm ad}/\Gamma_0=-0.85-\eta-1.75\beta+7.9\zeta/\gamma.
\end{equation}
If the gas passing through the horseshoe turn cools rapidly enough, the co-rotational torque is best approximated by the isothermal model
\begin{equation}
\Gamma_{\rm iso}/\Gamma_0=-0.85-\eta-0.9\beta.
\end{equation}
The scale of both torques is set by 
\begin{equation}
 \Gamma_0 = \left(\frac{q}{h}\right)^{2}\varSigma r^4 \varOmega^2,
\label{eq:scale-torque}
\end{equation}
where $q$ is the ratio of the planet mass to the central 
stellar mass, is the distance from the planet to the central body. 
This prescription is valid for low mass planets in unsaturated regions. 

If the gas can radiate away the surplus heat and restore the
temperature quickly, then the gas exerting the horseshoe drag should
be treated isothermally. The effective torque is interpolated between
the isothermal and adiabatic torque models. The ratio of the radiative
cooling time, $t_{\rm rad}$, to the dynamical time, $t_{\rm dyn}$,
defines the interpolation between the two regimes. We estimate 
\begin{equation}
t_{\rm  rad} \sim E/\dot{E}, 
\end{equation} 
where $\dot{E}$ is the rate of cooling by
radiation, or equivalently the divergence of the radiation
flux,$\v{F}$, and $E$ is the internal energy density of the gas,
$E=c_V\rho T$. For a cylinder of height $2H$ centered on the disk,
$\rho=\varSigma / 2H$, and the total energy within the cylinder is
\begin{equation}
 \int \! E\,{ \rm d} V = \int \! \frac{c_V \varSigma T}{2H} \,{\rm d} V=\pi r^2 c_V\varSigma T, 
\end{equation}
where the last step is true for a cylinder with a small radius with constant temperature and surface density throughout. Integrating the rate of cooling over the same volume gives
\begin{equation}
 \int \! \dot{E} \,{\rm d V} = \int \! \Div F{\rm dV}=\oint \!F{\rm dS} = 2 \pi r^2 \sigma T^4_{\rm eff}.
\end{equation}
This is true for small radii if the temperature outside of the cylinder is nearly the same as the temperature inside, since then the flux is approximately zero everywhere except the top and bottom of the cylinder. Substituting $T^4_{\rm eff}=T^4/\tau_{\rm eff}$, the radiation timescale is 
\begin{equation}
\frac{E}{\dot{E}}=\frac{c_V\varSigma \tau_{\rm eff}}{2\sigma T^3}.
\end{equation}
Defining $\Theta$ to be the ratio of the radiative timescale to the
dynamic timescale $2 \pi / \varOmega$, we have 
\begin{equation}
\Theta = \frac{c_V \varSigma \varOmega \tau_{\rm eff}}{4\pi \sigma T^3}.
\end{equation}
The effective torque acting on a planet is then given by interpolating
between the adiabatic and isothermal torques 
\begin{equation}
\Gamma_{\rm typeI} = \frac{\Gamma_{\rm ad}\Theta^2 + \Gamma_{\rm iso}}{\left(\Theta + 1\right)^2}.
\end{equation}
This interpolation, though not derived from physical principles, 
provides a smooth mathematical transition between the isothermal and 
adiabatic limits. We stress, though, that the precise nature of the 
transition has little influence on the results since, as shown  
in LPM10, most of the planet evolution in the disk through space and time 
occurs well within the adiabatic regime. The resulting torques 
are shown in \fig{fig:torqueProf}.

\subsection{Prescription for eccentricity and inclination damping}
\label{subsec:damping}
In addition to the torque that is responsible for Type~I migration, the 
gas disk also imparts a force on a planet that acts to dampen the planet's 
eccentricity $e$ and inclination $i$ \citep{Tanaka}. This dampening can be modeled 
as an exponential decay. Tanaka and Ward (2004) provide the
timescale 
\begin{equation}
 t_{\rm damp}=\frac{M_{\star}^2  h^4}{m_p\varSigma a_p^2 \varOmega}
\end{equation}
where $a_p$ is the semi-major axis of the planet, and the disk
  aspect ratio $h$ is given by equation~(\ref{eq:aspect}). \citet{Cresswell}
showed that the exponential decay is only applicable for small
inclinations and eccentricities, after which point the decay is best
modeled as a power law. Following their prescriptions, we use the
following timescales for the dampening of eccentricity and
inclination, respectively
\begin{equation}
  t_e = \frac{t_{\rm damp}}{0.780} \left(1-0.14\,\varepsilon^2+0.06\,\varepsilon^3 + 0.18\,\varepsilon\iota^2\right)
  \label{eq:ecc-time}
\end{equation}
\begin{equation}
   t_i = \frac{t_{\rm damp}}{0.544} \left(1-0.30\,\iota^2 + 0.24\,\iota^3 + 0.14\, \iota \varepsilon^2 \right)
 \label{eq:inc-time}
\end{equation}
where $\varepsilon$=$e/h$ and $\iota$=$i/h$.

These equations give a timescale for the damping that agrees well with
hydrodynamic simulations over the range of eccentricities and
inclinations observed in our simulations{\footnote{These timescales were derived for
isothermal disks. However, \citet{BK10,BK11} show that the behavior in radiative disks is
qualitatively similar. The quantitative difference is that the damping
timescale is slightly longer in radiative disks, because of the 
higher sound speed in an
adiabatic gas.  This is expected, since
the damping is provided by backreaction of waves generated in the disk
by an eccentric and/or inclined planet.}}. The timescales are
calculated at each step of the $N$-body integration, and the force
from the damping effect on a planet moving with velocity $v$ is
calculated as
\begin{equation}
\v{F}_{\rm damp,r}=-2 \frac{\left(\v{v} \cdot \v{r} \right) \v{r} }{r^2 t_e} m_p \hat{\v{r}}
 \label{eq:damp-force-r}
\end{equation}
\begin{equation}
\v{F}_{\rm damp,z}= -\frac{v_z}{t_i}m_p \hat{\v{z}},
 \label{eq:damp-force-z}
\end{equation} 
where $\hat{\v{r}}$ and $\hat{\v{z}}$ are the unit vectors in 
the $r$ and $z$ directions, respectively.

\begin{figure}
  \begin{center}
    \resizebox{\columnwidth}{!}{\includegraphics{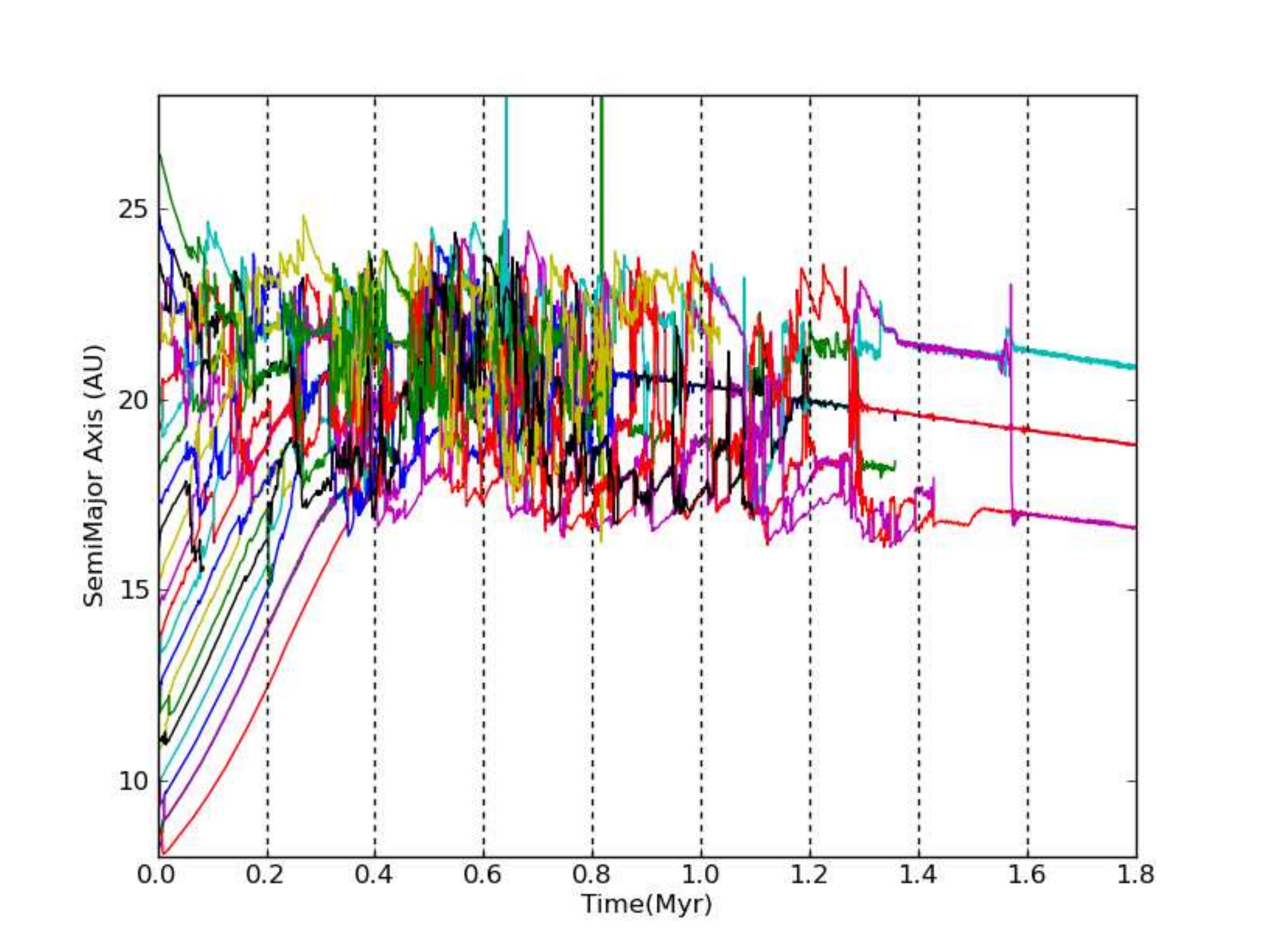}}
  \end{center}
\caption[]{The history of our fiducial model (H1), initially consisting of 23
planetary embryos of 0.75$M_{\earth}$ each. The planets migrate
towards the convergence zone. Chaotic $N$-body
interactions then start, and continue until a massive core is formed. The planetary core eventually reaches a mass of 13$M_{\earth}$ and continues to occupy the zero-torque region, with smaller bodies interior and exterior to it being forced into mean-morion resonances.}
\label{fig:fiducialHistory}
\end{figure}

\begin{figure}
  \begin{center}
    \resizebox{\columnwidth}{!}{\includegraphics{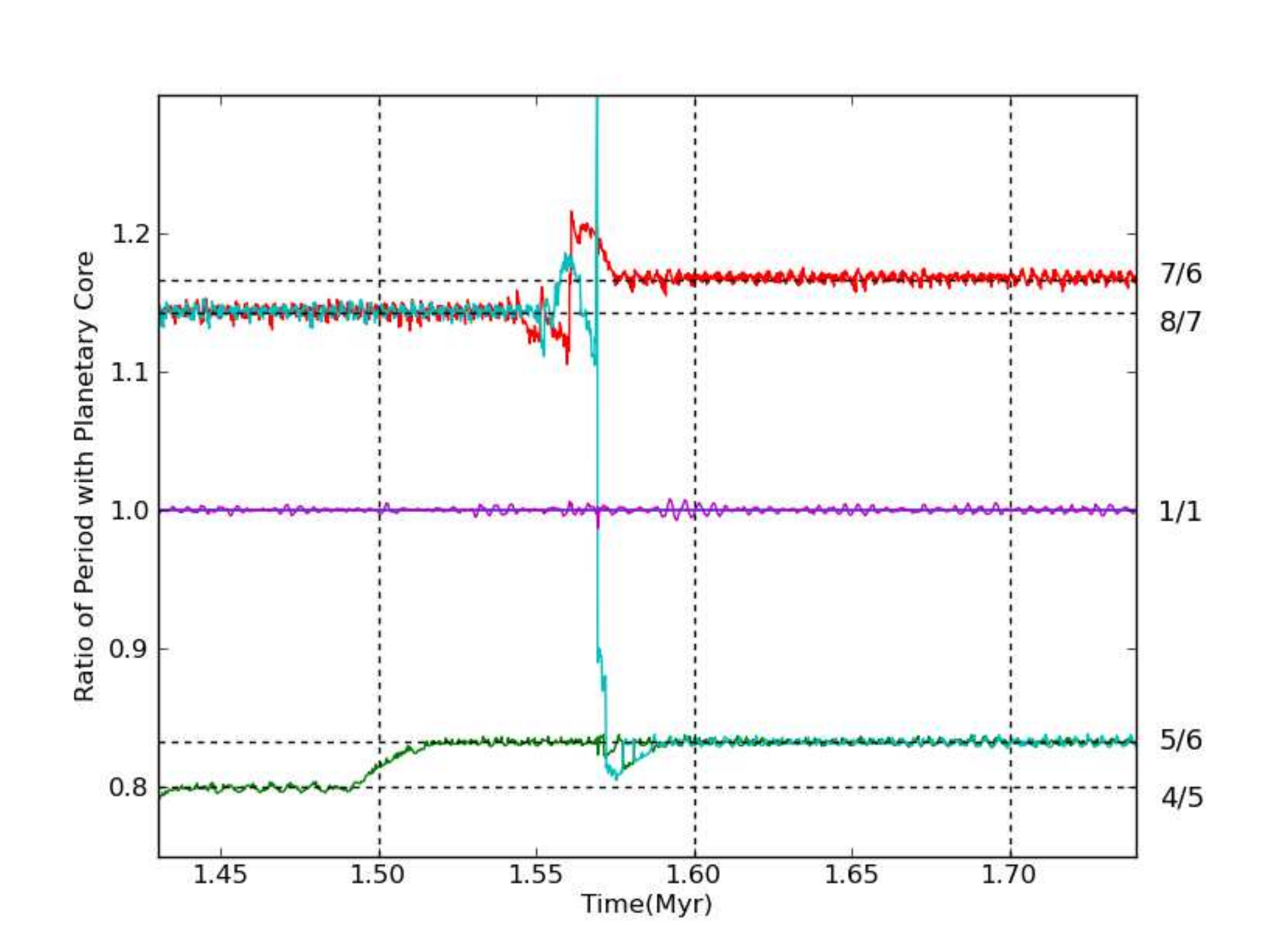}}
  \end{center}
\caption[]{The fiducial model, (H1), creates a large core that eventually reaches a mass of 13$M_{\earth}$. The remaining bodies are trapped in resonances with the planetary core. Following the transition at 1.57 Myr, the resonances remain stable throughout the rest of the simulation, with two 3$M_{\earth}$ bodies trapped in the 5:6 resonance, a 6$M_{\earth}$ body in a 7:6 resonance, and a 3$M_{\earth}$ embryo trapped as a Trojan of the planetary core. Similar results are seen in other runs.}
\label{fig:resonance}
\end{figure}

\begin{figure}
 \begin{center}
  \resizebox{\columnwidth}{!}{\includegraphics{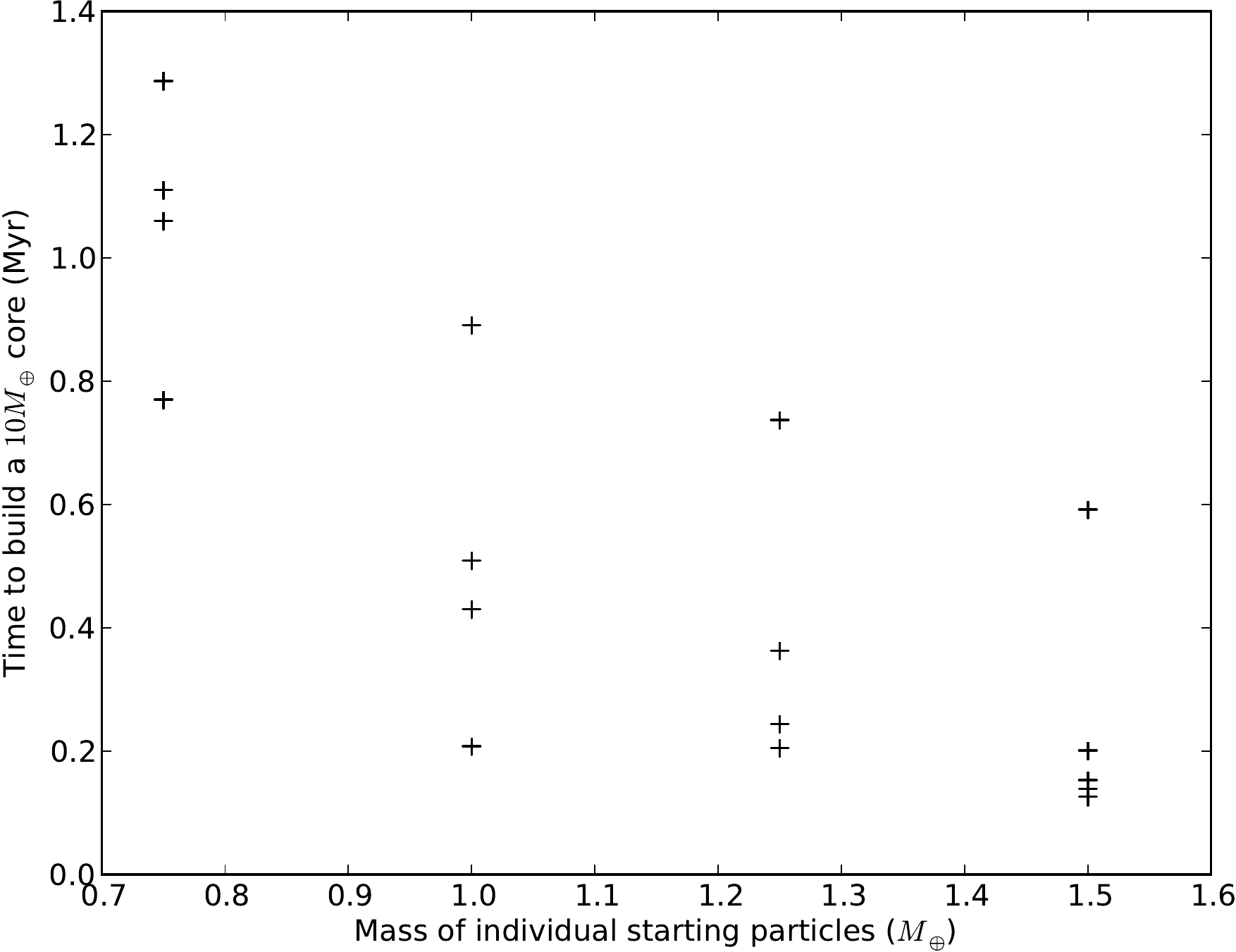}}
 \end{center}
 \caption[]{The time required to form a core of at least 10 $M_{\earth}$ 
   is shown to be a function of the initial masses of the planetary 
   embryos. These runs use a uniform mass for all bodies, whose 
   value sets the $x$ coordinate. All initial conditions except those 
   for the 1.5$M_{\earth}$ embryos have one simulation of the 5 runs
   fail to form a 10 $M_{\earth}$ core. These models are not included 
   in the figure.}
\label{fig:assemblyTime}
\end{figure} 

\begin{figure*}
  \begin{center}
    \resizebox{.8\textwidth}{!}{\includegraphics{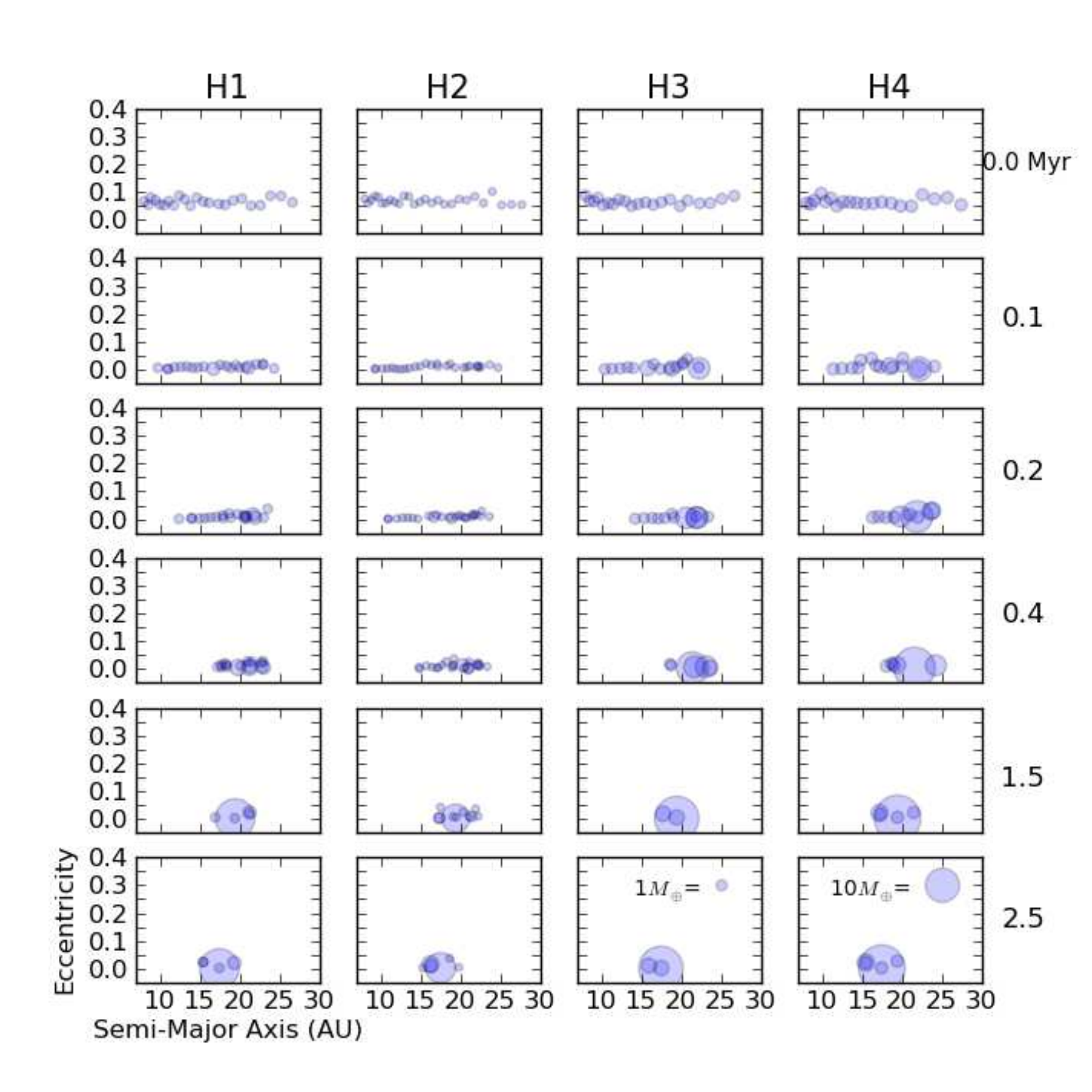}}
  \end{center}
 \caption[]{Time histories of models H, which initially consist 
of planetary embryos with 0.75, 0.5, 1.0 and 1.25 $M_{\earth}$, 
respectively. Each row depicts all four runs at a given time in its 
history--the top row represents the initial conditions, followed 
by 0.1, 0.2, 0.4, 1.5, and 2.5 Myr. Each panel depicts the semi-major 
axis and eccentricity of the bodies, with the area of a body scaling 
linearly with mass. Circles corresponding to masses
of 1\mearth 
and 10\mearth are shown for comparison in the bottom row.
All but the 0.5 $M_{\earth}$ run H2 form a core of at least 
10 $M_{\earth}$ by 1.5 Myr}
 \label{fig:masses}
\end{figure*}

\begin{figure*}
  \begin{center}
  \resizebox{.8\textwidth}{!}{\includegraphics{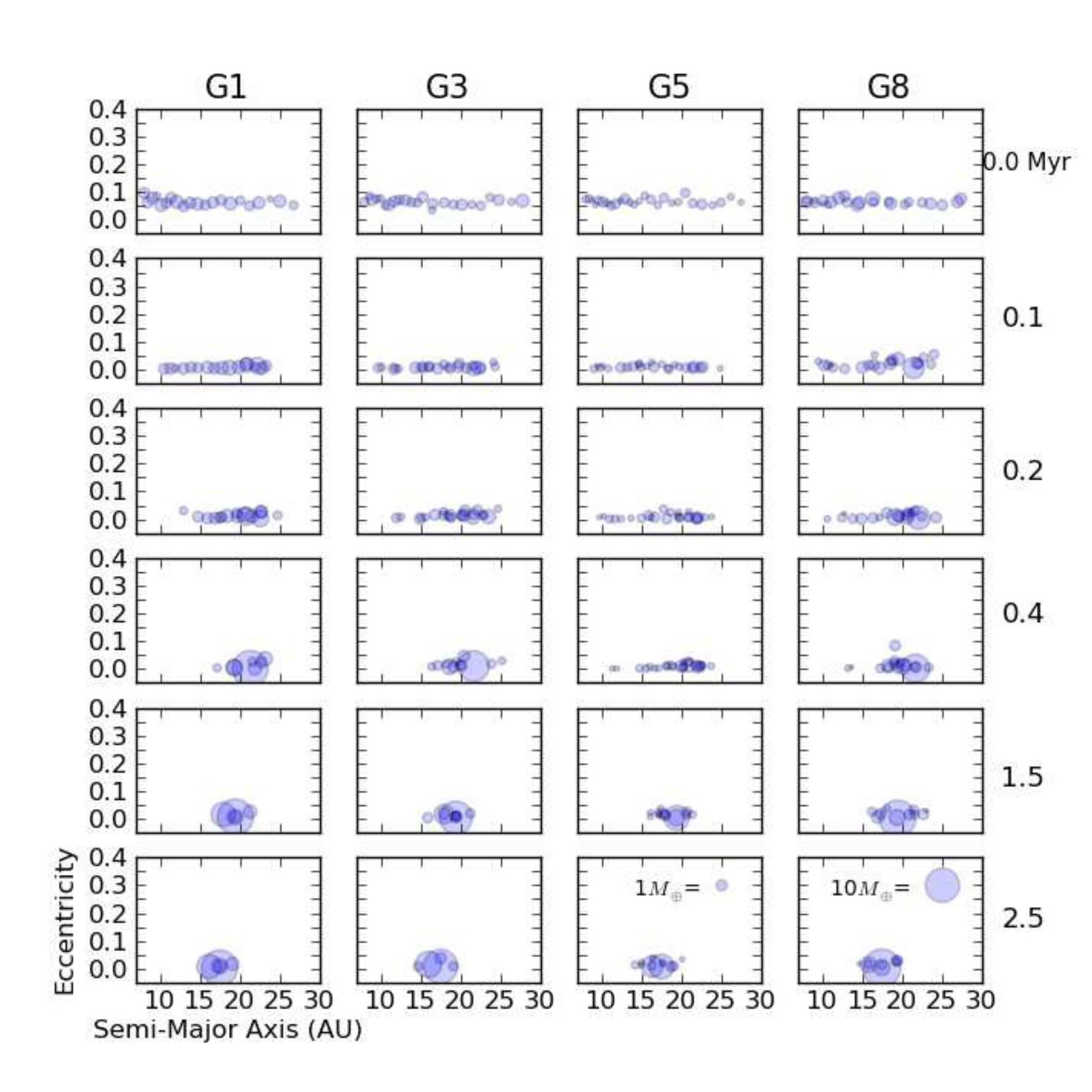}}
  \end{center}
\caption[]{Time histories of runs with a Gaussian mass 
distribution centered at 1.0$M_{\earth}$ (model G1) , 0.75$M_{\earth}$ (G3), and 0.5$M_{\earth}$ (G5) 
are presented, along with a superimposed distribution of the low density populations G2 \& G4 (G8). 
The top row shows the initial conditions, followed by the system after 0.1, 0.2, 0.4, 1.5 and 
2.5 Myrs. The results of core formation, and the time required to 
build a core, are comparable for the Gaussian and uniform 
mass distributions.}
\label{fig:randomHistory}
\end{figure*} 

\begin{figure*}
 \begin{center}
  \resizebox{\columnwidth}{!}{\includegraphics{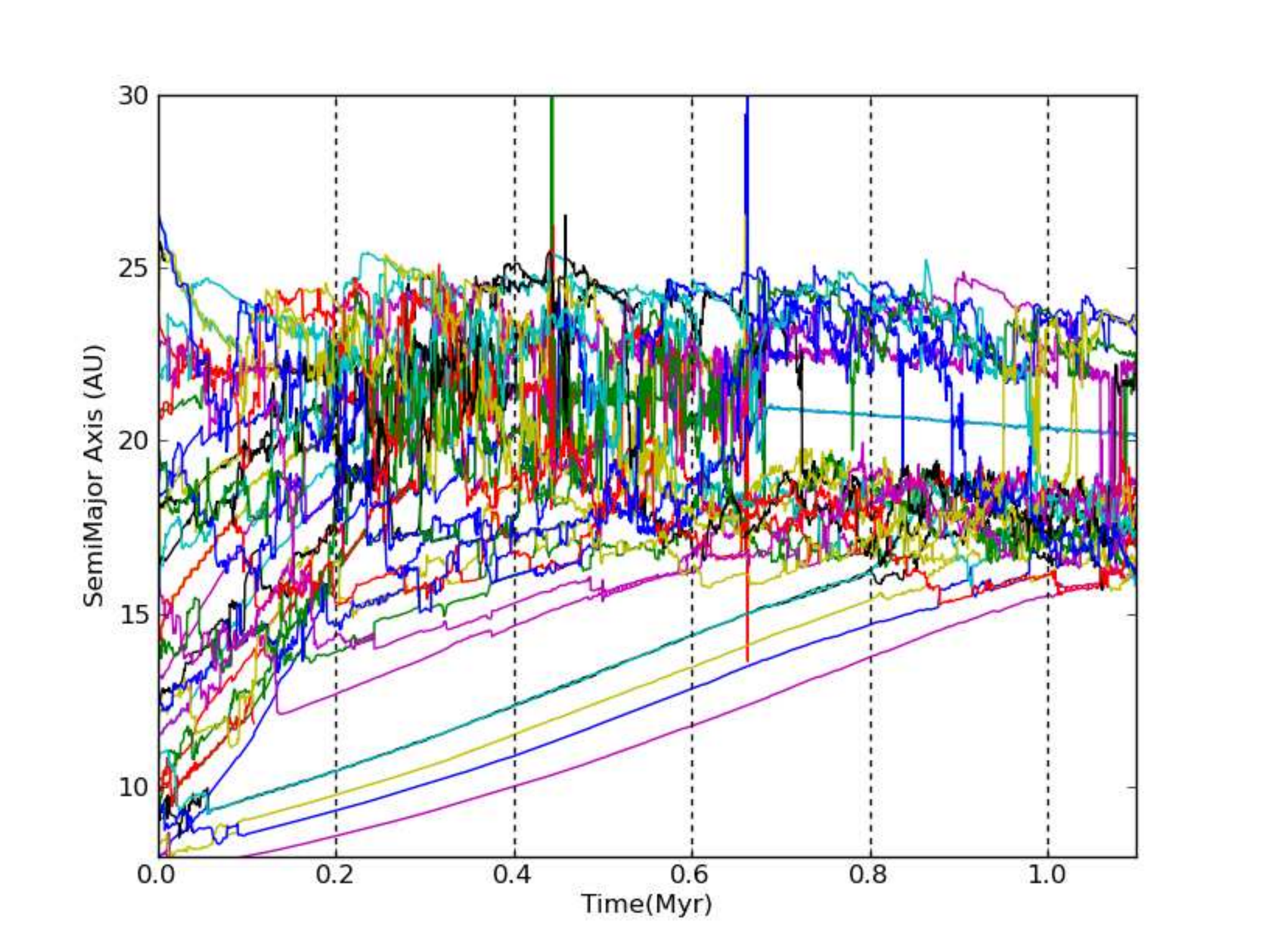}}
  \resizebox{\columnwidth}{!}{\includegraphics{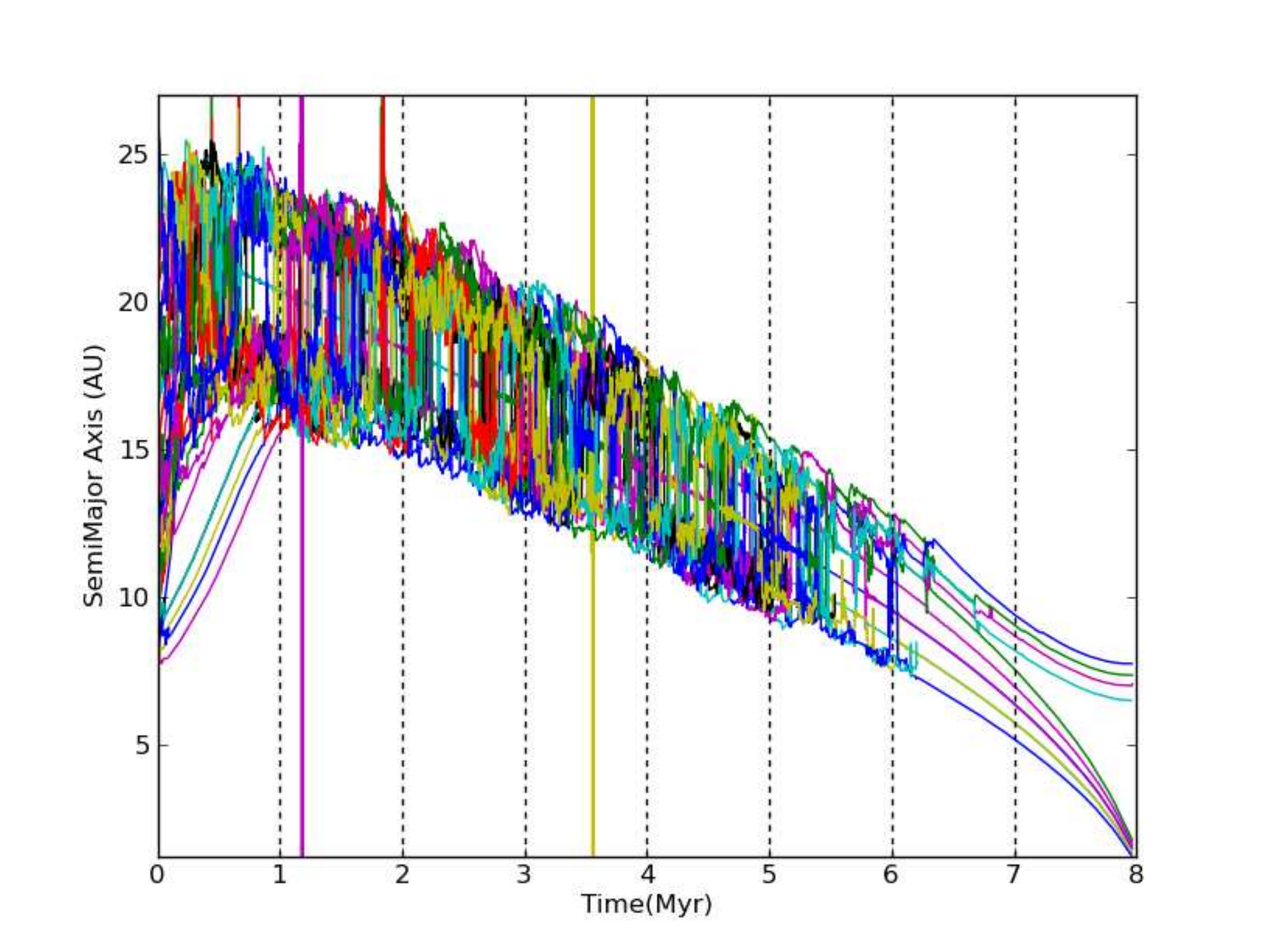}}
 \end{center}
\caption[]{{\it Left panel}. The formation history of 
model Sw3, which has 11 1$M_{\earth}$ bodies, migrating 
through a population of 25 0.25$M_{\earth}$ bodies. The
massive
bodies do not merge with the lower-mass ones as
the massive bodies rapidly 
migrate towards the convergence zone. Low mass
bodies are briefly  
scattered, but then resume their slower migration rate. The
massive bodies finish migrating to the convergence
zone by 0.3 Myr, while the low-mass bodies continue to
migrate until 1.0 Myr. An 8.5  
$M_{\earth}$ core is formed, though only two of the 25 
0.25$M_{\earth}$ bodies are consumed in forming it. {\it Right panel}. 
Extended history of the model. The swarm of smaller particles is seen 
to be long lived compared to the time required to form a planetary 
core. The swarm of low mass embryos does eventually merge with other 
bodies as they migrate inwards.}
\label{fig:bground}
\end{figure*}

\begin{figure*}
 \begin{center}
  \resizebox{.8\textwidth}{!}{\includegraphics{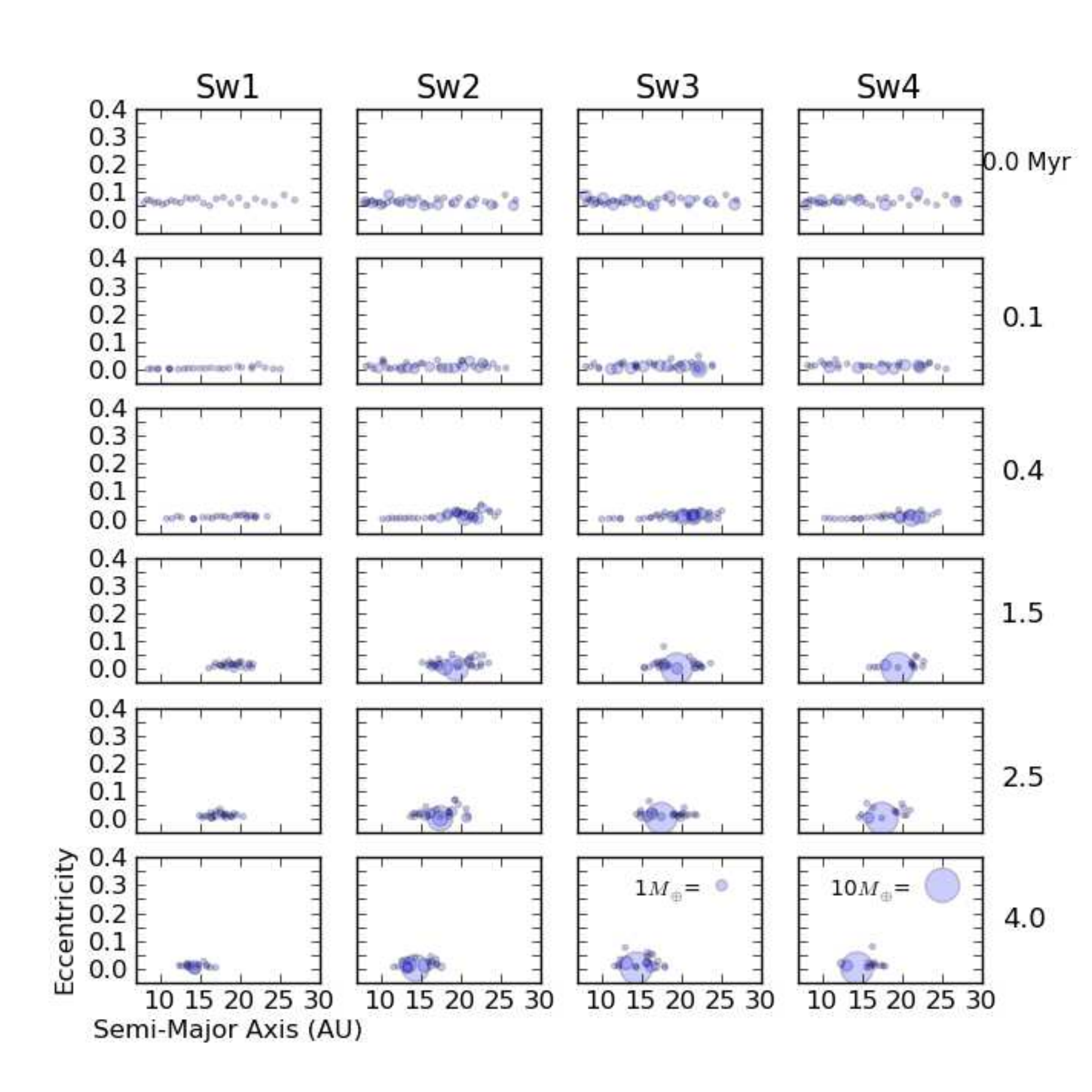}}
 \end{center}
\caption[]{Simulation histories of models Sw, where a population 
of planets evolves while interacting with a swarm of 25 low-mass (0.25 $M_{\earth}$) 
embryos. Column 1 shows the evolution of the swarm alone (model Sw1).
The more massive bodies quickly migrate to the convergence zone where 
they begin to form a planetary core. Only a small number 
of the 0.25$M_{\earth}$ bodies collide with anything, while most follow chaotic orbits 
in two distinct clusters, one interior and one exterior to the planetary core. 
These swarms eventually merge at later times, as seen in \fig{fig:bground}.}
\label{fig:historyBground}
\end{figure*}

\subsection{Turbulence}
\label{subsec:turbulence}

Small enough planets embedded in a turbulent disk have 
been shown to undergo random 
walks, rather than migrate inward in classical Type~I migration 
\citep{Nelson-Pap,Nelson}. The random walks can be 
modeled as a first order Markov process \citep{Rein} where the root mean 
squared torque and the autocorrelation time can be calculated up to a
constant by dimensional analysis. However, the correlation 
function of such a first order Markov process over the plane of the disk will 
be zero for all non-zero distances---so two planets in close proximity may 
feel completely different forces from this prescription for turbulence.

We use a model developed by \citet{Laughlin}
and further modified by \citet{Ogihara} that
treats the turbulent force as the gradient of a potential multiplied
by a scaling factor. In this model, the turbulent force
is continuous over the disk and correlated torques act on
planets near to each other. Due to the chaotic nature of $N$-body
problems, turbulence can lead to collisions that would not have
occurred otherwise, but during the close encounter of two bodies, the
difference between the turbulent forces acting on each body goes to
zero, and the relative velocity at collision is close to the velocity
for two bodies on the same orbit in the absence of
turbulence. Additionally, our results show that neighboring orbits
near the convergence zone can be stable when separated by as little as
0.4 AU at a distance of 11 AU from the central star, so the same
turbulent perturbations could affect both orbits. 

The turbulent potential, $\varPhi$, is
described as the sum of a large number of independent, scaled oscillation modes
\begin{equation}
  \varPhi_{c,m} = \psi r^2 \varOmega^2 \Lambda_{c,m}
  \label{eq:turbulent-potential}
\end{equation}
where $\Lambda_{c,m}$ is one oscillation mode, defined as
\begin{equation}
  \Lambda_{c,m} = \xi  e^{-(r-r_c)^2/\sigma^2} \cos\left(m\theta - \phi_c - \varOmega_c \tilde{t}\right) \sin\left(\pi \frac{\tilde{t}}{\Delta t}\right) 
  \label{eq:lambda}
\end{equation}

Each mode is defined by the azimuthal wavenumber $m$, as well as 
$r_c$ and $\phi_c$, which specify the initial center of the mode. The 
magnitude of the perturbation is set by $\xi$, whose value is set by a 
normal distribution with standard deviation of 1 and a mean value of 0. 
The planet's location is given by the radial coordinates $r$ and $\theta$, 
and the displacement above the disk, $z$, is assumed to be small enough to 
have a negligible effect. The mode comes into existence at time $t_0$ and 
evolves as a function of $\tilde{t} \equiv t_0 + t$. The lifetime of the 
perturbation is $\Delta t = 2\pi r_c / (m c_s)$, which is the sound-crossing 
time for a mode, where $c_s$ is the speed of sound. In \eq{eq:lambda}, 
$\varOmega_c$ is the Keplerian angular velocity at $r_c$.

The value for $m$ is chosen from a log-random distribution ranging from 1 
to 64. The value of $r_c$ is selected from a uniform random distribution 
ranging from $r_{\rm int}$ to $r_{\rm ext}$, 
and $\phi_c$ is selected from a uniform distribution from 0 to 2$\pi$. The 
radial profile of the perturbation is a Gaussian whose size is set by 
$\sigma=\pi r_c / (4m)$. At the beginning of a simulation, $n$ independent 
modes are generated, and as one mode expires another is generated so that at 
any time there are $n$ perturbations. This parameter $n$ is set to 50 in 
\citet{Laughlin} as well as in \citet{Ogihara}, 
but since we are investigating a disk that is an order of magnitude larger 
than that in \citet{Laughlin}, we have included 100 modes. 

In \eq{eq:turbulent-potential}, the magnitude of the potential used to model 
turbulence is determined by the dimensionless parameter $\psi$. This sets 
the scale for the magnitude of the turbulent force in comparison to the 
laminar forces due to Lindblad resonances and horseshoe
torques. \cite{Ogihara} suggest that $\psi$ can be anywhere from $\ttimes{-3}$ to $\ttimes{-1}$, depending on the strength of the instability driving the 
turbulence. The relation between $\psi$ and the \citet{Shakura} 
$\alpha$ parameter was explored in \cite{Baruteau}. 
\citet{Baruteau} found that 
\begin{equation}
 \psi \approx 8.5 \times 10^{-2} h \sqrt{\alpha}
\end{equation}
where the $h$ term enters due to the mode 
lifetime being set by the speed of sound. Our model does not have a
constant value for $h$, but we set this to 0.05 in this relation. 
We use two disks in our simulations, with $\alpha$ values of $10^{-2}$ and 
$10^{-3}$, so following \citet{Baruteau}, 
we take $\psi$
$4.25\times 10^{-4}$ and $1.34\times10^{-4}$, respectively. 

The sum of the $n$ independent oscillations, $\varPhi_{c,m}$, 
specifies the shape of the potential, $\varPhi=\sum\varPhi_{c,m}$, that, 
in \citet{Laughlin}, acted on the gas to  
generate density perturbations, which in turn applied a force to 
the planet. We do not apply the potential to our gas disk, which 
is only evolved in one dimension. Instead, the potential is used to
calculate the force acting on the planet directly. \citet{Ogihara} 
state that the force from such a turbulent disk can be expressed as 
\begin{equation}
  \v{F}_{\rm turb} = -C \grad{\varPhi}
  \label{eq:turbulent-force}
\end{equation}
where $C$ relates the size of the force acting from the 
potential onto the gas to the size of the force acting from the gas onto 
the planet
\begin{equation}
  C = \frac{64 \varSigma r^2}{\pi^2 M_\star}
  \label{eq:gamma}
\end{equation}

The strength of the force scales with the surface density of the gas at 
the planet's location, $\varSigma(r)$. \Eq{eq:turbulent-force} can be used 
to calculate the force due to turbulent perturbations on a planet 
at position ($r$,$\theta$) at time $t$. \citet{Ogihara} 
showed that all modes with wavenumbers higher than 6 can be left out
of the  
summation to determine the potential $\varPhi$, and so we follow this 
simplification, including only perturbations where $1 \leq m \leq 6$ 
in our calculation of $\varPhi$. 

\subsection{$N$-Body Code}

We simulate the behavior of multiple planetary embryos 
approaching a convergence zone, using the Bulirsch-Stoer $N$-body
code described by \citet{Sandor}, modified to
include the additional forces from the gas disk on each body. The
temperature and surface density profiles from the one-dimensional 
hydro simulations conducted previously are
read into the $N$-body simulation at each timestep and used in the
prescriptions described in Sections~\ref{subsec:torque}, \ref{subsec:damping},
and~\ref{subsec:turbulence} to determine the additional forces.

The total force resulting from Type~I migration,
damping, and turbulence
are computed for each body at the begining of a Bulirsch-Stoer
timestep, but are held constant in each refinement of the modified
midpoint algorithm during a given Bulirsch-Stoer timestep. The
Bulirsch-Stoer timesteps are a small fraction of the dynamical time of
the bodies, and are reduced during close encounters, so the
simplification of holding these forces constant during a timestep does
not change the overall behavior of the bodies. The net force acting on 
a body is given by
\begin{equation}
  \v{F}_{\rm total}=\v{F}_{\rm nbody} +\v{F}_{\rm typeI} +\v{F}_{\rm damp} +\v{F}_{\rm turb}
\end{equation}
where $F_{\rm nbody}$ represents the gravitational forces from the other bodies and the central star. The net force from the migration torques is
\begin{equation}
\v{F}_{\rm typeI}=\frac{\Gamma_{\rm typeI}}{r} \hat{\v{\theta}},
\end{equation}
while the components of 
$F_{\rm damp}$ are given by \Eqs{eq:damp-force-r}{eq:damp-force-z}, 
and $F_{\rm turb}$ is derived in \Eq{eq:turbulent-force}. 

A collision occurs if two bodies pass within a distance equal 
to the sum of the radii of the two bodies, where 
the radii are derived by assuming that all bodies have a bulk density of 
$2$ g\,cm$^{-3}$. Collisions result in the formation 
of one body with a mass and momentum equal to the sum of the two coliding 
bodies. This approximation is appropriate for bodies with the
  masses of planetary embryos that have escape velocities
  substantially exceeding typical collision velocities.

\section{Initial Conditions}
\label{sect:ic}

\subsection{Gas Disk Parameters}

The evolution of the gas disk was described in
\sect{sect:non-iso-disk}. The parameters that set the initial
profile of disk density and temperature are
$\alpha$, which sets the viscosity, the background temperature, $T_b$,
and the mass accretion rate, $\dot{M}$. MHD simulations consistently
suggest that $\alpha$ has a value of $10^{-2}$ \citep[e.g.][]{Davis},
though if the migration is taking place in a dead zone, the viscosity
should be smaller, though non-zero \citep{Stone,Oishi} 
Our simulations use a disk with $\alpha = 10^{-2}$, an accretion rate
$\dot{M}=10^{-7} M_{\sun}$~yr$^{-1}$, an external radius $r_{ext} =
30$~AU, and $T_b=10 K$. The parameters for the gas disks used in our
simulation are set so that the convergence zone lies within the
simulation at all times.

The photoionized wind has an escape radius $r_g = 5$~AU for a 1~$M_{\sun}$
star, and a photoevaporation rate set by a mass loss rate from the disk of
$\dot{M}_{\rm w} = 3 \times 10^{-8} M_{\sun}$~yr$^{-1}$. The accretion rate
$\dot{M}$, along with the photoevaporation 
rate $\dot{M}_{\rm w}$ and radius $r_g$, determines the lifetime of the
disk. Using the given values
for our parameters yields a 0.08 $M_{\sun}$ disk with an 8 Myr
lifetime.

\subsection{Planet Embryos}

The location of the convergence zone, where $F_{\rm typeI} \to  0$,
is determined by an opacity transition that 
in turn is set by the background temperature and the viscosity of the disk. 
In a disk with $T_b=10\,K$ and
  $\alpha=\ttimes{-2}$, the convergence 
zone begins near 22 AU and moves inwards 
towards the star on the viscous evolution timescale. Since this is the region of 
interest, and the computation time using the Bulirsch-Stoer method
scales as 
the number of bodies squared, we distribute the initial semi-major orbital 
radii of planet embryos uniformly around the radius of the convergence zone. Most runs 
begin with planetary embryos in the region between 8 and 28 AU. 

A consensus has not been reached on the
  initial mass function of planetary embryos. 
  Starting with an initial population of
  3000 equal mass (10$^{23}$\,g) seed planetesimals, and assuming a
  growth process dominated by binary collisions in the absence of gas
  drag and fragmentation, \citet{Kokubo}
  find a resulting mass
  distribution of power-law index $2.5\pm0.4$. The result of the
  opposite process, a collisional cascade from fragmentation of a
  parent body or collection of parent bodies with subsequent
  self-similar grinding of the fragments, yields a quasi-steady mass
  distribution of power law 11/6 
  \citep{Dohnanyi,Tanaka1996}. \citet{Lyra08} 
  found from hydrodynamical models of 
  protoplanetary disks including interacting decimeter-sized particles
  and gas tides, that the resulting mass distribution of lunar and
  Mars-sized embryos obeyed a power-law index of $2.3\pm0.2$,
  consistent with coagulation \citep{Kokubo} and, at 3$\sigma$, with 
  fragmentation \citep{Dohnanyi}, suggesting a combination of both processes.
  However, none of these power laws match the observed mass distribution
  of the Asteroid and Kuiper belts (e.g., \citealt{Cuzzi,Morbidelli09}), 
  which are characterized by multiple power-laws (see fig 1 of 
  \citealt{Morbidelli09}). Instead of adopting these power-laws, we use 
  simple mass distributions, detailed below. We justify this by noting that
  the relaxed mass distribution is nearly independent of the initial 
  mass distribution \citep{Kokubo}. The parameters of the
various models we ran are shown in \tab{table:runs}. We detail them in
this and in the next section.

First, in our fiducial model (model H1 in \tab{table:runs}), 
the initial masses of the planetary embryos are all 
set to 0.75 $M_{\earth}$. The innermost embryo is placed in a
circular orbit with radius $a_1 = 8$~AU, and each 
succesive embryo is placed so that neighbors~$i$ and~$j$ are initially 
separated by $N \times R_{mH}$, where the mutual Hill radius
\begin{equation}
  R_{mH}=\left(\frac{m_i+m_j}{3M_{\star}}\right)^{1/3}\left(\frac{a_i+a_j}{2}\right),
\end{equation}
and $a_j=a_i+N\,R_{mH}$. In the fiducial model H1, we use $N=3$.
Several other runs were performed with varying masses of initial bodies, but maintaining the 
constraint of spacing consecutive bodies by three
mutual Hill radii. These runs are all
labeled {\it H}, for ``Hill spacing'', in \tab{table:runs}.

Second, we used a Gaussian
distribution of the initial masses of planet
embryos. This is done by setting the mean mass and standard
deviation for the distribution, with a floor of 0.1 $M_{\earth}$ and a
ceiling of $2 M_{\earth}$.  These models are labeled {\it G}, for
``Gaussian'', in \tab{table:runs}. In these runs, the mutual Hill
radii constraint is still used to determine initial
locations, but the 
number of Hill radii between neighboring bodies is varied
between runs to model
low density and high density populations. Three further simulations are
performed with an initial population of bodies resulting from
superimposing two low density populations (models G7--G9).

Third, we ran initial conditions consisting of two populations of
planetary embryos with different masses, to investigate how
massive, quickly migrating bodies interact with
lower-mass bodies that migrate on much longer
timescales. The low-mass population in this model consists of 25
embryos with a mass of 0.25 $M_{\earth}$ separated by 4 mutual Hill
radii, while the massive population consists of planets with masses of
0.75 or 1.0 $M_{\earth}$ separated by either 6 or 10 mutual Hill
radii. The two populations are generated independently of one another,
so a large and small body may initially be in close proximity to one
another. These models are labeled {\it Sw}, for ``swarm'', in
\tab{table:runs}.

Fourth, we use an
initial configuration that scales with the amount of disk mass within the
orbital radius of a planetary embryo. The mass of the planetary
embryos is held constant, so the initial spacing between the semimajor
axes of neighboring embryos is varied so that the mass of the gas
in the annulus separating the two bodies is constant. Assuming
that the dust-to-gas ratio in the disk is 0.01, and
planetesimal formation turns $15\%$ of the dust into
planetary embryos, then the initial spacing between two neighboring
bodies 
\begin{equation}
  \Delta{r}=(6.67 \times 10^2)\frac{M_{p}}{2 \pi r \varSigma} \:.
  \label{eq:annulus}
\end{equation}
where the numerical factor is 
the inverse of the product of the dust-to-gas ratio 0.01 and the 
planet formation efficiency 0.15. These models are labeled 
{\it SD}, for ``surface density'', in
\tab{table:runs}. One simulation is performed using a Gaussian mass
distribution, spaced using the SD criteria. This simulation is
included in the SD suite of simulations. 

The initial eccentricities and inclinations of planets are also selected 
randomly from Gaussian distributions. The inclination is the absolute value 
of a Gaussian distribution with 0 mean and $0.05^{\circ}$ 
as the standard deviation.The mean value for the initial eccentricity 
is 0.05, with a standard deviation of 0.02. If the eccentricity is negative, 
a new value is generated until the value is non-negative. Due to the magnitude 
of the damping force, initial eccentricities and inclinations are found to be 
transient and do not play a significant role in the simulations.

\section{Results}
\label{sect:results}
\subsection{The Fiducial Model}
The history of the fiducial model (H1), with 23 bodies each having a
mass of 0.75$M_{\earth}$ and separated 
by $3 R_{mH}$, is shown in \fig{fig:fiducialHistory}. Initially, 
the motions of the embryos are dominated by the Type~I torque, leading to 
rapid convergence. The convergence leads to close encounters between 
neighboring bodies, resulting in collisions and orbit swapping. The close 
encounters and chaotic scattering and collisions persist until the number of 
remaining planets is small enough to quiescently migrate
inwards.

The sharp peaks near 600,000 and 800,000 years are not 
planets being scattered out to large radii, but rather the result of solving 
for the orbital elements, assuming the planet is orbiting the central body, 
when in reality the bodies have formed a transient binary system. Such binary 
systems and satellite capture have been observed in other $N$-body 
simulations that include the effects of a gas disk, such as 
\cite{Cresswell}. 

The fiducial model has a total initial mass of 17.25 $M_{\earth}$
in planet embryos. In it, a $10.5
M_{\earth}$ core forms at the zero-torque orbit in 1.29 Myr. No further collisions happen after this, as the remaining 
bodies are trapped in mean-motion resonances with the massive planetary 
core (\fig{fig:resonance}). Two $3 M_{\earth}$ bodies are trapped in the 5:6 
resonance with the planetary core, and a $6 M_{\earth}$ body is trapped outside 
the orbit of the planetary core, in a 7:6 resonance with it. An additional 
$3 M_{\earth}$ embryo orbits within the co-orbital region of the planetary core 
as a Trojan. Other simulations show similar behavior, with accretion eventually
being halted by resonance trapping once the planetary core is massive enough. 

\begin{figure}
 \begin{center}
  \resizebox{\columnwidth}{!}{\includegraphics{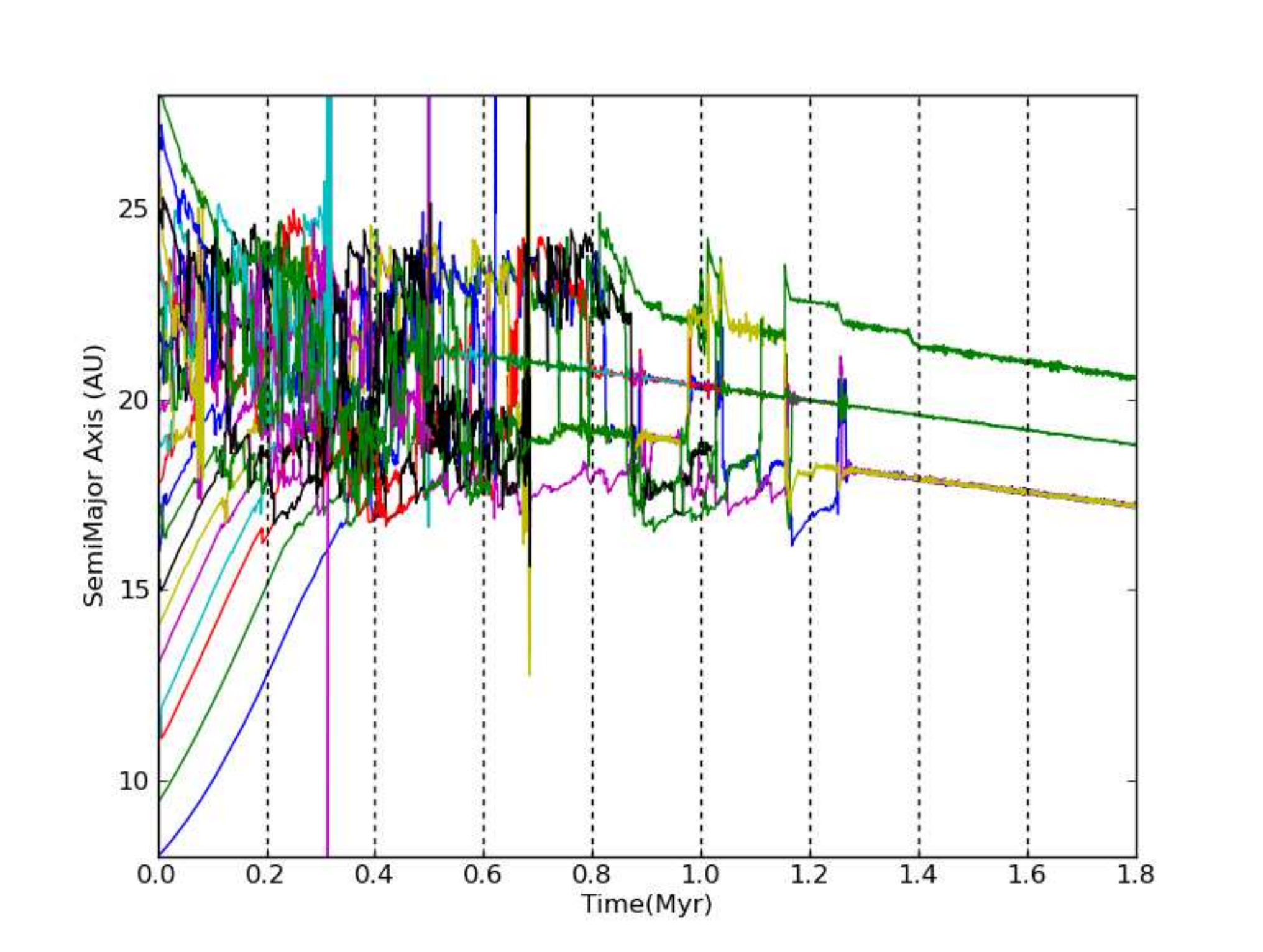}}
 \end{center}
\caption[]{The history of model SD1, a run with 
  0.75$M_{\earth}$ planets 
  initially distributed such that the surface 
  density of planets is 0.15\% of the gas surface density. The results 
  are comparable to the fiducial model, suggesting that mass distribution 
  does not alter the qualitative description we have given of the formation 
  of planetary cores.}
\label{fig:sdFiducial}
\end{figure}

\subsection{Variations of Planetary Embryo Mass}

The fiducial model begins with planetary embryos of 0.75 $M_{\earth}$. We also
simulate disks populated with embryos of 0.5, 1.0, 1.25, and 1.5 
$M_{\earth}$ separated, as in the fiducial case, by $3 R_{mH}$. These 
correspond to models H2-H5 in \tab{table:runs}.
Given the dependence of the Hill radius on the mass of the planet, 
systems consisting of more massive embryos contain 
a greater total mass. 

Since the magnitude of the migration torque, \eq{eq:scale-torque}, varies 
as the square of the mass of the planet, and thus the acceleration varies 
linearly with mass, lighter embryos take longer to migrate to the 
convergence zone. This, combined with gravitational focusing, explains 
why the time required to build a core is a strong function of the initial 
mass of the embryos (\fig{fig:assemblyTime}). The fiducial model built a 10$M_{\earth}$ core in 1.28 
Myr, while the run using 1.0$M_{\earth}$ embryos reached a 10$M_{\earth}$ core 
in 0.43 Myr. The runs with 1.25 and 1.5 $M_{\earth}$ embryos formed cores in 
0.21 and 0.15 Myr, respectively. The 0.5 $M_{\earth}$ embryos failed to produce 
a 10 $M_{\earth}$ body, but did form an 8 $M_{\earth}$ core after 2.22 Myr.
 
To see if these results are characteristic of such a mass distribution, 
identical simulations to the ones above with initial planet masses of 
0.75, 1.0, 1.25, and 1.5 $M_{\earth}$ are run, though the azimuthal 
distribution of the planets is varied, and so the chaotic interactions lead 
to different results. The dependence of the time required to assemble a 
massive core does hold for the five runs of each initial mass. There is, 
however, a wide upper bound for the time to form a core, so much so that each 
distribution other than the 1.5 $M_{\earth}$ runs (model H5) had one 
run of the 5 fail to form a 10 $M_{\earth}$ core. These are not plotted in 
\fig{fig:assemblyTime}, but the resulting systems were interesting, often 
consisting of two or more massive bodies coorbiting with one another stably 
for the rest of the 8 Myr simulation. 

The formation history of runs H1--4 are shown in \fig{fig:masses}. 
The simulations with more massive embryos are 
seen to rapidly develop large planets, while the simulations 
consisting of embryos with initial masses 
less than $1 M_{\earth}$ remain dispersed 
after $4\times 10^5$~yr. The simulations
demonstrate that, while the time  
required to form a core can vary by over an order of magnitude for embryos 
whose initial mass differs by a factor of two, 
the resulting core and the arrangement of the remaining bodies is 
similar for all the simulations. The results 
presented in \fig{fig:masses} are representative of the collection of 
simulations run with these initial conditions.

Runs G1--9 also suggest that the rate of core
formation is highly dependent on the mass  
of the embryos. Two runs with a mean mass of $1.0 M_{\earth}$ and a standard 
deviation of $0.3M_{\earth}$ (G1 and G2) built large cores within 0.3 Myr. The initial 
masses for the two distributions are generated independently, which also 
varies the radial distribution, since planets are initially 
separated by $3 R_{mH}$. The core
formation history of one  
of these runs is shown in \fig{fig:randomHistory}, along with the formation histories 
of G3, G5, and G8. The results of these runs 
suggest that the trend seen in \fig{fig:assemblyTime} can be used to estimate 
the time required to assemble a core from a Gaussian mass distribution with a  given mean mass.

If the mean of the initial mass distribution is lowered to 0.5$M_{\earth}$ 
(models G5 and G6) the time required to migrate bodies to the convergence zone increases 
significantly, and when cores are built, they are smaller and take longer to 
form. The mass distribution is again Gaussian with a standard deviation of 
0.2$M_{\earth}$, though there are lower and upper mass limits of 
0.1$M_{\earth}$ and 2.0$M_{\earth}$, respectively. Two runs were performed, 
with an independent mass distribution generated for each. The first run took 
2.5 Myr to build a 6.9$M_{\earth}$ core, and is depicted in 
\fig{fig:randomHistory} beside the history of a run composed of only 
0.5$M_{\earth}$ planetary embryos. The second run took 3.2 Myr to form 
a 9.3$M_{\earth}$ core. 

The runs with a population of 0.25$M_{\earth}$ bodies interspersed
with more massive embryos did not result in the smaller bodies being
swept up in collisions with the rapidly migrating large
embryos. Instead, the larger embryos scatter the 0.25$M_{\earth}$
objects into high eccentricity orbits and continue to migrate towards
the convergence zone where mergers between large bodies go on to
create cores. This can be seen in \fig{fig:bground} (left panel) by
the two characteristic timescales for migration, which, aside from a
few brief scattering events, occur more or less independently of one
another. Following the formation of a core with mass
$\sim5M_{\earth}$, smaller bodies are seen to collide with the core or
become trapped as Trojans. Many, though, avoid close encounters with
the planetary core and continue to interact chaotically with the other
remaining small planetary embryos in two swarms, one interior and one
exterior to the planetary core. Neglecting the (substantial)
  effect of gas accretion onto the planetary core, the model shows
  that these swarms eventually merge at later times,
as seen in the right panel of \fig{fig:bground}, 
which depicts the whole 8\,Myr of the
simulation. \Figure{fig:historyBground} depicts the evolution of the
background swarm of 0.25$M_{\earth}$ by themselves (model Sw1) as well
as three cases where more massive bodies are interspersed into the low
mass swarm (models Sw2-Sw4). The mass of the larger bodies, as well as
the number of larger bodies, is not seen to have a noticeable effect
on the migration and merging of the smaller bodies.

\subsection{Mass Distribution}

An alternative to determining the initial positions of planetary
embryos by requiring them to be a set number of mutual Hill radii from
their neighbor is to require that the distribution of mass in the form
of planets mirror the mass distribution of the gas disk. This leads to
a larger number of planets starting at radii exterior to the
zero-torque orbit. In \fig{fig:sdFiducial}, we compare the
history of fiducial run H1 
with run SD1, which differs from the fiducial run only in that the
planetary mass is linearly proportional to the mass of gas within an
annulus, following \eq{eq:annulus}.

The early period of convergence is 
skewed to larger radii in the latter, but 
as can be seen in \fig{fig:torqueProf}, the torque exterior to the convergence 
zone is of the same magnitude as the torque interior to the convergence zone. 
The time to migrate from initial locations to close enough to the convergence 
zone that $N$-body effects take over is not significantly affected by altering 
the initial radial distribution of planets.

\section{Caveats}
\label{sect:caveats}

The prescription for Type~I migration used in this paper 
was derived by \cite{Paard2010} using 
2D simulations.  Those authors noted that the gas dynamics may 
act differently in 3D.

Our simulations assume that the Type~I migration torque derived for an
isolated body in a non-isothermal disk also applies to many bodies that
pass arbitrarily close to one another. This close proximity may
significantly alter the migration torque or lead to a more rapid
saturation of the torque. Such proximity is often shortlived, but in 
the case of a satellite being captured could lead to a change in the torque 
acting on the pair lasting thousands of years. A preliminary
two-dimensional hydrodynamical simulation of two bodies migrating in close 
proximity suggests that the wakes of the planets drive brief periods of
large variations in the torque. This will be explored in future work.

The initial convergence takes place while planets are separated from one 
another, so using the prescription for the torque on an isolated body is 
reasonable. Following convergence, close gravitational encounters dominate 
until a planetary core is formed. In several simulations, the resulting system 
had two large bodies that had a combined mass greater than 10$M_{\earth}$
but failed to produce a single massive core. Studying such a system in a 
simulation that accounts for gap opening by the planets
in the gas disk could shed light on how such an arrangement will evolve.

\citet{BK10} showed that the horseshoe torque is 
reduced as eccentricity is increased, and shuts down as the 
radial excursions associated with eccentricity exceed the 
width of the co-rotation region. 
\citet{BK11} further showed that outward migration can be 
sustained for inclinations up to about 4.5$^\circ$, after which migration  
stalls in general, owing to the lower densities away from the disk midplane.
We assess the effect of eccentricity and inclination in the next section.

In addition, the prescription for the co-rotational torque that 
we use considers only the unsaturated regime. We 
calculate for the parameters  
used in the present work that the effects of diffusion, explored 
in \citet{Paard2011} for alpha disks, imply that only larger planets, with masses 
exceeding $\approx 4 M_{\earth}$, will experience sustained outward 
migration. Our choice of working with the unsaturated torque stems from the 
results of \citet{Baruteau}, \citet{Uribe}, and 
\citet{Baruteau11}. These works measured the co-rotational torques in 
turbulent disks, finding that the stochastic turbulent fluctuations work 
toward keeping the co-rotational torque unsaturated even in locally isothermal 
simulations. \citet{Baruteau11} have conclusively shown that horseshoe 
dynamics exists in turbulent disks, and 
argue that it occurs when 
the amplitude of the U-turn drift rate exceeds that of the turbulent 
velocity fluctuations. This condition may or may not occur for smaller 
planets, for which the U-turn drift rate is reduced. Unfortunately, 
neither \citet{Baruteau11} nor \citet{Uribe} could numerically resolve the 
expected horseshoe region width in this mass range. We thus recognize 
saturation for smaller planets as a possibility, but regard the question as 
unsettled.

Our conclusions about resonance trapping neglect the effect of
  interactions with the wakes of the growing planetary core.  These
  may be sufficiently strong to knock lower-mass bodies out of the
  resonances that we find. Even if the resonances do remain, our
  assumption about evolving the gas disk independently of the planets
  breaks down once the planetary core has reached a mass near 10
  $M_{\earth}$.  At that point it begins first to accrete gas, and
  then, when its Hill radius exceeds a disk scale height, to open a
  gap in the disk. Our models account for neither development, and so
  should be taken as only qualitatively indicative (e.g. the right
  panel in Figure~\ref{fig:bground}) after any object reaches this
  mass. 

We note that the actual critical mass necessary for runaway 
  gas accretion is a sensitive function of the luminosity of the forming 
  planet, which is determined by the release of gravitational binding energy 
  of the accreted material---usually assumed to be provided steadily 
  by planetesimals. In our model, this continuous luminosity is
  essentially zero, which would lead to a very small critical core mass \citep{Papaloizou}. The 
  energy is instead provided in discrete violent collisions. A model that incorporates a
  self-consistent treatment of gas accretion and gap formation by a growing planet 
  with envelope luminosity provided by both planetesimal accretion and binary collisions 
  would be required to fully assess the initial stages of giant planet formation.

\subsection{Comparison to Recent Work}
\label{sect:hn12}

\begin{figure}
 \begin{center}
  \resizebox{\columnwidth}{!}{\includegraphics{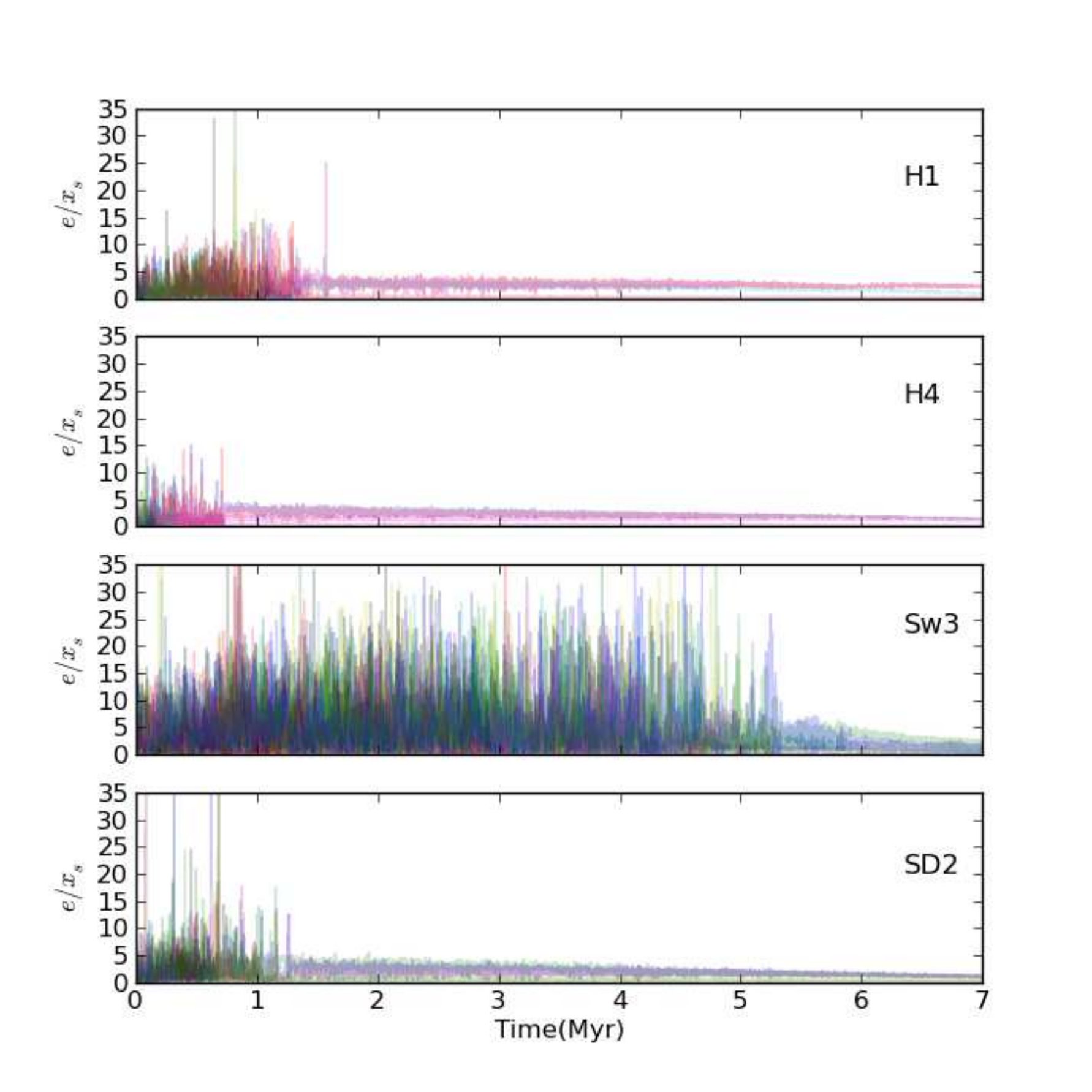}}
 \end{center}
\caption[]{The eccentricity $e$ normalized by the dimensionless half-width of the 
co-rotational region $x_s$, for the protoplanets in four of our models. The eccentricity 
is pumped well above the critical value of $e/x_s=1$ associated with shutdown of the 
horseshoe drag (see text).}
\label{fig:eccentricities}
\end{figure}

\begin{figure}
 \begin{center}
  \resizebox{\columnwidth}{!}{\includegraphics{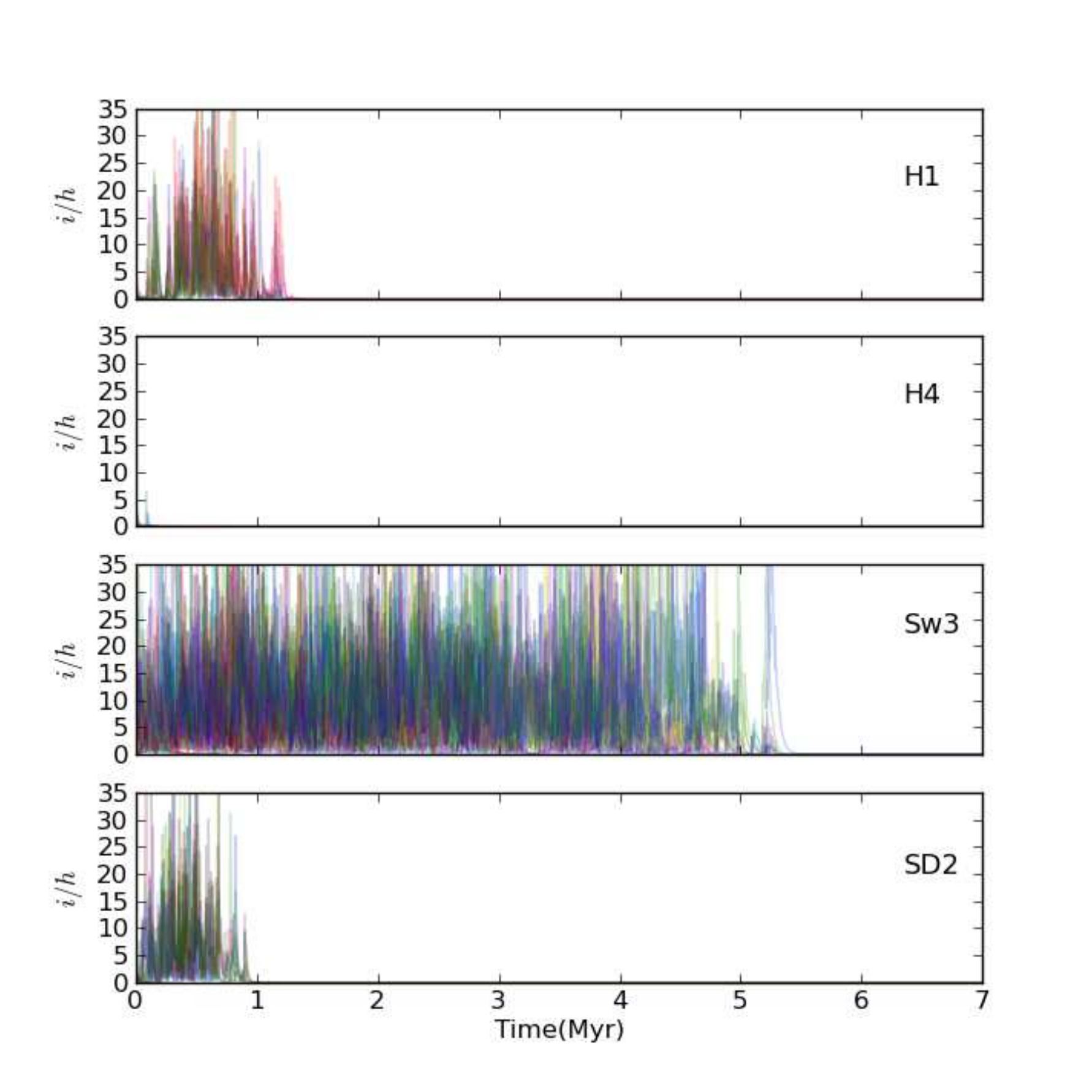}}
 \end{center}
\caption[]{The inclinations $i$ normalized by the aspect ratio $h$ of 
the disk, for the same models as in \fig{fig:eccentricities}. The inclinations 
are pumped well above the critical value of $i_{\rm crit}=0.08$ (i.e., 
$i_{\rm crit}/h$ = 1.6) beyond which migration stalls (see text). This is a 
potentially important phenomenon, perhaps countering the shutdown of 
co-rotational torques due to finite eccentricity shown in 
\fig{fig:eccentricities}.}
\label{fig:inclinations}
\end{figure}

The work presented in this paper examines the same
regime as that studied by HN12, which was made public after the
initial submission of our work. Although both 
works deal with the impact of co-rotation torques and convergent migration 
on the growth of protoplanets, there are differences in the models that 
are worth highlighting, in order to understand the different results 
obtained. 

First, we discuss the gas evolution
model. In the present work we model the  
gas density by equating the radial velocity with the viscous inflow, the 
azimuthal velocity with the Keplerian value, and solving
the resulting diffusion equation  
in the presence of photoevaporation. The initial condition for density is set by defining a 
global value for mass accretion rate and viscosity ($\alpha$) value, which in turn 
define the surface density \citep{Papaloizou}. Having defined the 
density, the disk is assumed to maintain thermal equilibrium between heating by 
viscous dissipation and cooling via black body radiation in a background 
radiative field of temperature $T_b$, according to equation~(\ref{eq:temperature}). In HN12, the 
density and temperatures are set as global power-laws (that may drop
well below the physical value of $T_b \simeq 10$~K set by the surrounding
molecular cloud, and even below the temperature of the cosmic
microwave background). The 
viscous-photoevaporative evolution of the density is mimicked by an exponential 
decay with a fixed e-folding time $\tau_{\rm disk}$; the temperature is evolved 
dynamically assuming thermal diffusion in the radial direction. 

Next we consider the growth of the
protoplanets. While our work concerns the formation of 
a core of a giant planet in the vicinity of a convergent migration zone via binary 
collisions, HN12 attempt a more comprehensive approach. In addition to
collisions with planets of terrestrial size, 
their protoplanets also accrete from a disk of 10\,km planetesimals subject to Stokes drag. 
They also include an approximate scheme for gas accretion based on fits to the fast 1D 
giant planet formation model of \citet{Movshovitz}. In their
  model, gas capture is a function only 
of core mass and time, starting at roughly 3~ $M_{\oplus}$, with runaway gas
accretion setting in at 30~$M_{\oplus}$. The gas
accretion is artificially cut off at either the Jovian mass or at the 
gas isolation mass, whichever is lower.  

A minor difference between our work and HN12 lies in the inclination 
and eccentricity damping. Both works use prescriptions that are based on the analytical 
formulae derived by \citet{Tanaka2002} and \citet{Tanaka}. HN12 use the same inclination damping, 
but a smoother version for eccentricity damping. We use the model of \citet{Cresswell} that improves 
on \citet{Tanaka} by including empirical fits to the transition from exponential to quadratic damping 
at higher eccentricities and inclinations. 

Finally we consider the prescription for
migration of the protoplanets. HN12 model
saturation effects, while we chose to work with the fully 
unsaturated torque, as explained in the previous subsection. Another major difference is our 
inclusion of stochastic migration \citep{Nelson-Pap,Nelson} 
caused by the inclusion 
of turbulent jitter (\sect{subsec:turbulence}). 

As a result, HN12 find a bimodal formation process. Cores that grow too massive too fast 
migrate into the star, whereas cores of moderate mass slowly migrate outward to large distances, 
accreting gas and becoming giant planets at wide orbits, mostly beyond 10\,AU. Super-Earths 
or Neptune-mass planets were not easily formed. 

The difference in the results that we find are readily understood in terms of these differences in the modeling. 
First, without a background radiation field of temperature $T_b$ below which the disk cannot cool, 
the model of HN12 lacks the transition to isothermality that gives rise to a limit of outward migration 
in our models, usually between 10--20\,AU, depending on the initial parameters used. 
Second, their inclusion of saturation in HN12 
introduces a mass-dependency to both the
location and existence of convergence zones. 
Rapid inward migration occurs for planets outside of a narrow range in mass. Lastly, our inclusion 
of turbulent jitter dynamically excites the resonant convoy, leading to more gravitational interaction 
than seen in HN12, and faster core growth. These differences highlight the fact that planet migration 
and growth depend very sensitively on the disk model adopted.

In two out of their suite of forty simulations, HN12 apply a eccentricity cutoff to the co-rotational 
torque, taking in to account the findings of \citet{BK10}. 
They find that the eccentricity excursions 
of the protoplanets easily exceed the threshold of $e/x_s = 1$, where $x_s=1.1/\gamma^{1/4} \sqrt{q/h}$ 
is the dimensionless half-width of the co-rotational zone, therefore quenching the co-rotational torques. 
We measure $e/x_s$ in four of our models. The result is shown in 
\fig{fig:eccentricities}. We see that the eccentricies excited by the N-body interactions and 
the turbulent jitter on the protoplanets far exceeds the stringent threshold of $e/x_s = 1$, confirming 
the result of HN12. However, the same processes that account for eccentricity pumping should 
also lead to inclination growth. As shown by \citet{BK11}, migration stalls in general when 
the inclination rises above a threshold $i_{\rm crit}\approx 4.5^\circ \approx 0.08$. We plot in 
\fig{fig:inclinations} $i/h$ for the same models as in \fig{fig:eccentricities}. A comparison of the 
two figures suggests that eccentricity and inclination are indeed correlated. The aspect ratio being 
$h=0.05$, we have $i_{\rm crit}/h = 1.6$. Figure~\ref{fig:inclinations} shows that the inclinations grow 
well beyond this modest threshold. This indicates that although the high eccentricities would lead to a 
shutdown of the co-rotational torques and resuming of fast inward migration, the associated high 
inclinations would mitigate the effect by also shutting down the Lindblad torque and, consequently, 
migration in general.

\section{Conclusions}
\label{sect:conclusions}

We have studied the evolution of multiple planets migrating in
a non-isothermal gas disk, assuming unsaturated migration torques
\citep{Paard2010}. Initially, the planets migrate towards the
convergence zone as was described by LPM10 for individual
planets in such a disk. Once the planets are near the convergence
zone, however, the close proximity to other planets leads to chaotic
interactions and collisions, eventually resulting in the formation of
several large bodies that often merge to form a single massive
planetary core. The planetary core dominates the dynamics of the 
remnant bodies, and the smaller bodies tend to get trapped in
mean-motion resonances with the core. Some become trapped as Trojans
with the core, and the whole system migrates inward on the
  viscous timescale as the disk
evolves, maintaining the resonances. As the goal of this investigation
was to study planetary embryos undergoing Type~I migration, the
simulation does not model accretion onto the planetary core or Type~II
migration. Both should be treated in future work.

Since the 
migration torque scales with the mass of the planet, simulations that 
initially consist of larger mass planetary embryos produce planetary cores 
rapidly, while simulations that start with less massive bodies often take more 
than 1~Myr to form a core. The rate of core formation does not appear to 
depend on the initial distribution of planets in the disk, but does depend 
on the mass of the planetary embryos that coalesce to form the core. A 
realistic protoplanetary disk would consist of planetary embryos with a range 
of masses, but we have shown that even in the case of a Gaussian
  distribution of embryo masses the trend of large embryos migrating faster and merging 
earlier holds. 

The location of the convergence zone is determined by the
temperature profile of the gas disk. This, in turn, is determined by
the turbulent viscosity as parameterized by $\alpha$, the total
disk mass, and background illumination
from the central star and neighboring stars. The convergence zone falls at
larger radii for larger values of $\alpha$
and the disk temperature. More massive stars have
hotter disks, leading to convergence zones at larger
radii. Direct imaging has indeed revealed a population of giant
planets at radii of order 100~AU
\citep[e.g.][]{Marois08,Kalas08,Oppenheimer08,Marois10} around
stars significantly more massive than the Sun. Our results
offer a possible means of building gas giant planets
via fast core accretion at such large radii.

\acknowledgments {Partial support for this work comes from NASA grant
  NNX07AI74G and NSF grant
AST10-09802. Some of the computations were performed at the AMNH
Parallel Computing Center.\\}


\begin{thebibliography}{}
\expandafter\ifx\csname natexlab\endcsname\relax\def\natexlab#1{#1}\fi
\bibitem[{{Alibert et al.}(2004)}]{Alibert} Alibert, Y.; Mordasini, C.; \& Benz, W. 2004, A\&A, 417, 25
\bibitem[{{Balbus \& Hawley}(1991)}]{Balbus91} Balbus,S. A., \& Hawley, J. F., 1991, ApJ, 376, 214
\bibitem[{{Baruteau et al.}(2011)}]{Baruteau11} Baruteau, C., Fromang, S., Nelson, R. P., \& Masset, F. 2011, A\&A, 533, A84
\bibitem[{{Baruteau \& Lin}(2010)}]{Baruteau} Baruteau, C., \& Lin, D. N. C. 2010, ApJ, 709, 759
\bibitem[{{Baruteau \& Masset}(2008)}]{Baruteau-Masset} Baruteau, C., \& Masset, F. 2008, ApJ, 672, 1054
\bibitem[{{Bate et al.}(2003)}]{Bate} Bate, M. R., Lubow, S. H., Oglive, G. I., \& Miller, K. A. 2003, MNRAS, 341, 213
\bibitem[{{Bell et al.}(1997)}]{Bell} Bell, K. R., Cassen, P. M., Klahr, H. H., \& Henning, Th. 1997, ApJ, 486, 372
\bibitem[{{Bitsch \& Kley}(2011)}]{BK11} Bitsch, B., \& Kley, W. 2011, A\&A, 530, A41
\bibitem[{{Bitsch \& Kley}(2010)}]{BK10} Bitsch, B., \& Kley, W. 2010, A\&A, 523, A30
\bibitem[{{Birnstiel et al.}(2009)}]{Birnstiel} Birnstiel, T., Dullemond, C. P., \& Brauer, F. 2009, A\&A, 503, L5
\bibitem[{{Chambers et al.}(1996)}]{Chambers96} Chambers, J. E., Wetherhill, G. W., \& Boss, A. P. 1996, Icarus, 119, 261
\bibitem[{{Cresswell \& Nelson}(2008)}]{Cresswell} Cresswell, P., \& Nelson, R. P. 2008, A\&A, 482, 677
\bibitem[{{Crida et al.}(2006)}]{Crida} Crida, A., Morbidelli, A., \& Masset, F. 2006, Icarus, 181, 587 
\bibitem[{{Cuzzi et al.}(2010)}]{Cuzzi} Cuzzi, J. N., Hogan, R. C., \& Bottke, W. F. 2010, Icarus, 208, 518
\bibitem[{{D'Angelo et al.}(2002)}]{D'Angelo} D'Angelo, G., Henning, T., \& Kley, W. 2002, A\&A, 385, 647
\bibitem[{{Davis, Stone \& Pessah}(2010)}]{Davis} Davis, S. W., Stone, J. M., \& Pessah, M. E. 2010, ApJ, 713, 52
\bibitem[{{Dohnanyi}(1969)}]{Dohnanyi} Dohnanyi, J. S. 1969, J. Geophys. Res. 74, 2531 
\bibitem[{{Fleming \& Stone}(2003)}]{Stone} Fleming, T., \& Stone, J.~M.\ 2003, \apj, 585, 908
\bibitem[{{Hellary \& Nelson}(2012)}]{Hellary} Hellary, P., \& Nelson, R. P. 2012, arXiv:1112.2997 (HN12)
\bibitem[{{Hollenbach et al.}(1994)}]{Hollenbach} Hollenbach, D., Johnstone, D.,Lizano, S., \& Shu, F. 1994, ApJ, 428, 654
\bibitem[{{Hubeny}(1990)}]{Hubeny} Hubeny, I. 1990, ApJ, 351, 632
\bibitem[{{Kalas et al.}(2008)}]{Kalas08} Kalas, P., Graham, J. R., Chiang, E., Fitzgerald, M. P., Clampin, M., Kite, E. S., Stapelfeldt, K., Marois, C., \& Krist, J. 2008, Science, 322, 1345
\bibitem[{{Kley \& Crida}(2008)}]{Kley08} Kley, W. \& Crida, A. 2008, A\&A, 487, L9
\bibitem[{{Kokubo \& Ida}(1996)}]{Kokubo} Kokubo, E. \& Ida, S. 1996, Icarus, 123, 180
\bibitem[{{Laughlin et al.}(1994)}]{Laughlin} Laughlin, G., Steinacker, A., \& Adams, F. C. 2004, ApJ, 608, 489
\bibitem[{{Lynden-Bell \& Pringle}(1974)}]{L-B} Lynden-Bell, D., \& Pringle, J. E. 1974, MNRAS, 168, 603
\bibitem[{{Lyra et al.}(2010)}]{Lyra} Lyra, W.,Paardekooper, S.-J.,\& Mac Low,  M.-M. 2010, ApJ, 715, L66 (LPM10)
\bibitem[{{Lyra et al.}(2008)}]{Lyra08} Lyra, W., Johansen, A., Klahr, H., \& Piskunov, N. 2008, A\&A, 491, L41
\bibitem[{{Marois et al.}(2010)}]{Marois10} Marois, C., Zuckerman B., Konopacky Q. M., Macintosh B., \& Barman T. 2010, Nature, 468, 1080
\bibitem[{{Marois et al.}(2008)}]{Marois08} Marois, C., Macintosh, B., Barman, T., Zuckerman, B., Song, I., Patience, J., Lafreni\`ere, D., \& Doyon, R. 2008, Science, 322, 1348
\bibitem[{{Morbidelli et al.}(2008)}]{Morbidelli08} Morbidelli, A., Crida, A., Masset, F., \& Nelson, R. 2008, A\&A, 478, 929
\bibitem[{{Morbidelli et al.}(2009)}]{Morbidelli09} Morbidelli, A., Bottke, W., Nesvorn\'y, D., Levison, H. 2009, Icarus, 558, 573
\bibitem[{{Movshovitz et al.}(2010)}]{Movshovitz} Movshovitz, N., Bodenheimer, P., Podolak, M., \& Lissauer, J. J. 2010, Icarus, 209, 616
\bibitem[{{Nakamoto \& Nakagawa}(1994)}]{Nakamoto} Nakamoto, T. \& Nakagawa, Y. 1994, ApJ, 421, 640
\bibitem[{{Nelson}(2005)}]{Nelson} Nelson, R. P. 2005, A\&A, 443, 1067 
\bibitem[{{Nelson \& Papaloizou}(2004)}] {Nelson-Pap} Nelson, R. P., \& Papaloizou, J. C. 2004, MNRAS, 350, 849
\bibitem[{{Ogihara et al.}(2007)}]{Ogihara} Ogihara M., Ida, S., \& Morbidelli, A. 2007, Icarus, 188, 522
\bibitem[{{Oishi and Mac Low}(2009)}]{Oishi} Oishi, J.~S., \& Mac Low, M.-M.\ 2009, \apj, 704, 1239
\bibitem[{{Oppenheimer et al.}(2008)}]{Oppenheimer08} Oppenheimer, B. R., Brenner, D., Hinkley, S., Zimmerman, N., Sivaramakrishnan, A., Soummer, R., Kuhn, J., Graham, J. R., Perrin, M., Lloyd, J. P., Roberts, L. C., Jr., \& Harrington, D. M. 2008, ApJ, 679, 1574
\bibitem[{{Paardekooper et al.}(2010)}]{Paard2010} Paardekooper, S.-J., Baruteau, C., Crida, A., \& Kley, W. 2010, MNRAS, 401, 1950
\bibitem[{{Paardekooper et al.}(2011)}]{Paard2011} Paardekooper, S.-J., Baruteau, C., \& Kley, W. 2011, MNRAS, 410, 293
\bibitem[{{Paardekooper \& Papaloizou}(2008)}]{Paard2008} Paardekooper, S.-J.,\& Papaloizou, J. C. B. 2008, A\&A, 485, 877
\bibitem[{{Paardekooper \& Papaloizou}(2009)}]{Paard2009} Paardekooper, S.-J.,\& Papaloizou, J. C. B. 2009, MNRAS, 394, 2283
\bibitem[{{Paardekooper \& Mellema}(2006)}]{Paard2006} Paardekooper, S.-J.,\& Mellema, G. 2006, A\&A, 459, L17
\bibitem[{{Papaloizou \& Terquem}(1999)}]{Papaloizou} Papaloizou, J. C. B., \& Terquem, C. 1999, ApJ, 521, 823
\bibitem[{{Pollack et al.}(1996)}]{Pollack} Pollack, J. B., Hubickyj, O., Bodenheimer, P., Lisaur, J. J., Podolak, M., \& Greenzweig, Y. 1996, Icarus, 124, 62
\bibitem[{{Rein \& Papaloizou}(2009)}]{Rein} Rein, H., \& Papaloizou, J. C. A\&A, 497, 595
\bibitem[{{S\'andor et al.}(2011)}]{Sandor} S\'andor, Zs., Lyra, W., \& Dullemond, C. 2011, ApJ, 728, L9
\bibitem[{{Shakura \& Sunyaev}(1973)}]{Shakura} Shakura, N. I., \& Sunyaev, R. A. 1973, A\&A, 24, 337
\bibitem[{{Tanaka et al.}(2002)}]{Tanaka2002} Tanaka, H., Takeuchi, T., \& Ward, W. R. 2002, ApJ, 565, 1257
\bibitem[{{Tanaka et al.}(1996)}]{Tanaka1996} Tanaka, H., Inaba, S. \& Nakazawa, K. 1996, Icarus, 123, 450
\bibitem[{{Tanaka \& Ward}(2004)}]{Tanaka} Tanaka, H., Ward, \& W. R. 2004, ApJ, 602, 388
\bibitem[{{Uribe et al.}(2011)}]{Uribe} Uribe, A., Klahr, H., Flock, M., \& Henning, Th. 2011, ApJ, 736, 85
\bibitem[{{Ward}(1997)}]{Ward} Ward, W. R. 1997, Icarus, 126, 261
\bibitem[{{Yang et al.}(2009)}]{Yang09} Yang, C.-C., Mac Low, M.-M., \& Menou, K. 2009, ApJ, 707, 1233
\bibitem[{{Yang et al.}(2011)}]{Yang11} Yang, C.-C., Mac Low, M.-M., \& Menou, K. 2011, in press (arXiv:1103.3268)

\end{thebibliography}
\end{document}